\documentstyle[proceedings,12pt,epsfig,colordvi]{JHEP}
\conference{Heavy Flavours 8, Southampton, UK, 1999}
\title{Experimental results on hadronic c decays}
\author{Brian Meadows \\University of Cincinnati\\ Department of Physics
                      \\Mail Location 11 \\ Cincinnati OH 45221
                      \\E791 Collaboration
                      \\ \email {Brian.Meadows@UC.edu}}
\abstract{Recent results involving hadronic decays of charmed mesons
 and baryons are reviewed.  Information relevant to decay mechanisms
 and to light quark scalar mesons is discussed.}
\dedicated{}

\def\PRref#1&#2&#3(#4){\unskip\ #1~\bf #2\rm, #3 (#4)}
\def\NPref#1&#2&(#3)#4{\unskip\ #1~\bf #2\rm (#3) #4}

\def\half{{1\over 2}}

\let\mathrm=\rm

\def\Xcp{\Xi_c^+}
\def\Lc{\Lambda_c^+}
\def\Xiz{\Xi^{\circ}}
\def\Xistz{\Xi^{\ast\circ}}

\def\Dz{D^{\circ}}

\def\Dp{D^+}

\def\Dsp{D_s^+}

\def\Dstr{D^{\ast}}

\def\Kp{K^+}
\def\Km{K^-}
\def\Kstz{K^{\ast\circ}}

\let\Kst=\Kstz
\let\Kstbar=\Kstzbar
\def\pip{\pi^+}
\def\pim{\pi^-}
\def\pim{\pi^-}

\def\rhoz{\rho^{\circ}}
\def\fz{f^{\circ}}

\def\etc{{\sl etc~}}

\def\half{{1\over 2}}

\begin{document}
\section{\bf Introduction}

New information on hadronic decays of charmed particles obtained in
the last year is reviewed.  Meson results are reviewed first, then
new baryon decay data and finally Cabbibo suppressed decays of both
mesons and baryons.

\def\rhoz{\rho^{\circ}}
\def\etp{\eta^{\prime}}
\def\mz{m_{\circ}}
\def\fz{f^{\circ}}
\def\f2{f_2}
\def\ubar{\overline{u}}
\def\dbar{\overline{d}}
\def\sbar{\overline{s}}
\def\bbar{\overline{b}}
\def\cbar{\overline{c}}
\def\tbar{\overline{t}}
\def\Aj{{\cal A_{\rm j}}}
\def\aj{a_{\rm j}}
\def\deltj{\delta_j}
\def\fj{f_{\rm j}}
% _________________________________________________________________________

\section{\bf New Results on Hadronic Decays of Mesons}

In general, the lowest order tree diagrams leading to such decays are shown in
figure \ref{fig-meson_trees}.
\FIGURE{%
\begin{minipage}{0.40\textwidth}
  \centering
  \epsfig{file=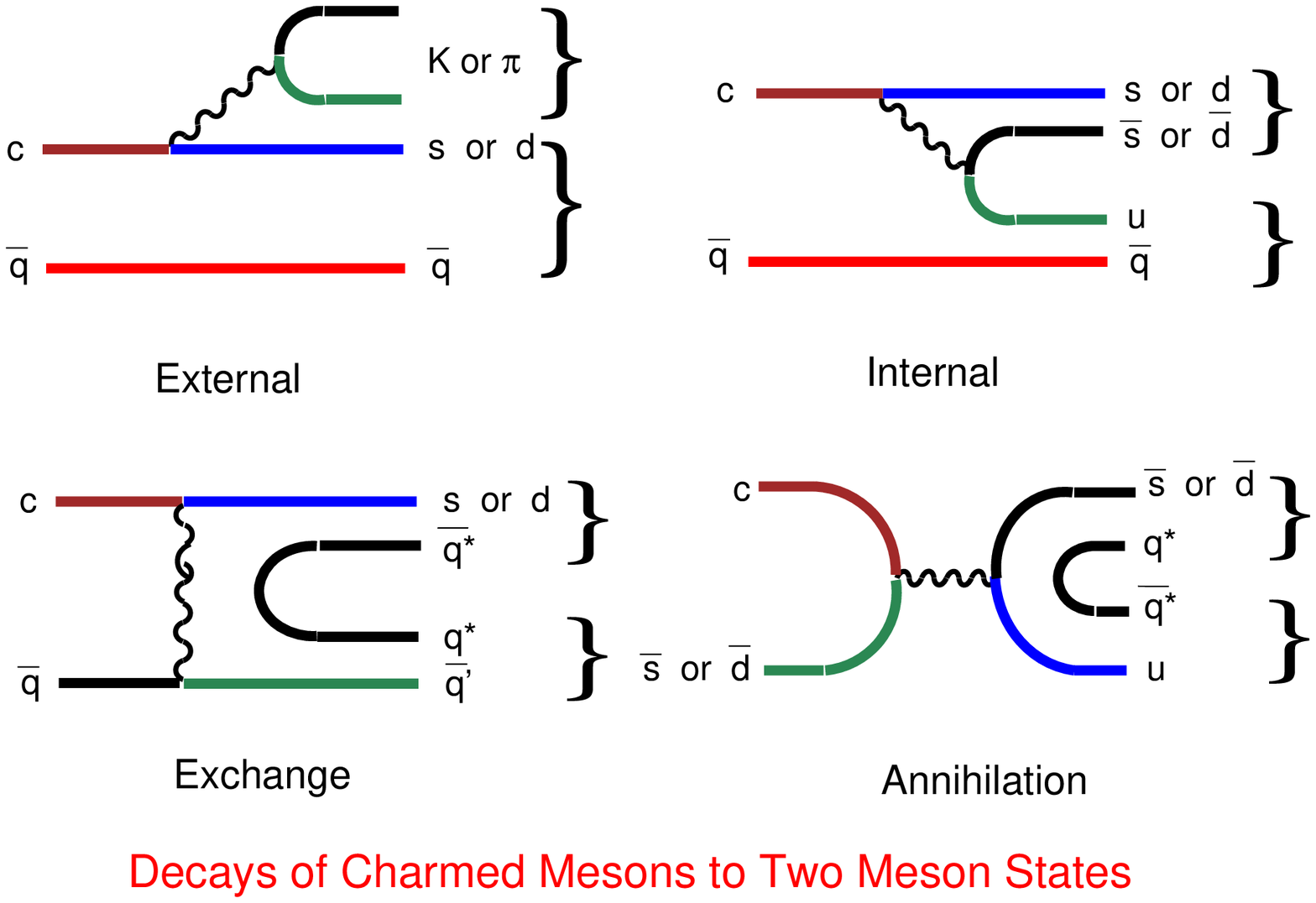,width=2.8in,angle=0}
  \caption{Lowest order tree diagrams for decays of charmed mesons to two meson
           final states.  Additional final state mesons can easily be included.}
  \label{fig-meson_trees}
\end{minipage}
}% \FIGURE
In a recent review \cite{Browder:1996af} experimental evidence for the relative
strengths of these diagrams is discussed.  The internal spectator process is
colour suppressed with respect to the external, and destructive interference
between the two, possible in the case of $\Dp$ but not in $\Dz$ nor
$\Dsp$, provides the most likely explanation for the lifetime
relationships  $\tau_{\Dz}\sim\tau_{\Dsp}\sim\tau_{\Dp}/2.5$.
$W$-exchange is helicity suppressed and little evidence for the annihilation
diagram has yet been seen.

\subsection{\bf New Data on Decays of $\Dp~(\Dsp)\rightarrow\pim\pip\pip$ + cc.}

These decay modes provide further information on this picture.  Decays of $\Dsp$
to $\rhoz\pip,~\fz\pip$, or $\pim\pip\pip$ could occur only via the annihilation
diagram, or by inelastic final state interaction (FSI).  Though no evidence for
annihilation had yet been seen, these decays are at least Cabbibo favoured.

Decays of $D$ mesons to scalar + pseudo-scalar final states
observed in earlier data from E691 \cite{Anjos:1989pu} and
from E687 \cite{Frabetti:1997sx} also provide information on scalar mesons.
Using samples of $236\pm 20 \Dp$ and $98\pm 12 \Dsp$ events
they found no evidence for $\Dsp$ decay to $\rhoz\pip$ or to $3\pi$, but they
found that $\Dp$ decay was dominated by both these modes.  In fact,
$\Dsp$ decay was dominated by quasi two body decay while $\Dp$ decay
was dominated by the 3 body mode.

E791 has new data \cite{E791-d3pi:1999ab} with higher statistical precision
from which some new conclusions can be drawn.  Important new data 
pertaining to low mass, light quark scalar meson states decaying to 
$\pip\pim$ were also found.

\subsubsection{Branching ratio}
E791's signals for the $D$ mesons decaying to three pions are shown in figure
\ref{fig-e7913pi}.
\FIGURE{%
 \begin{minipage}{0.40\textwidth}
  \epsfig{file=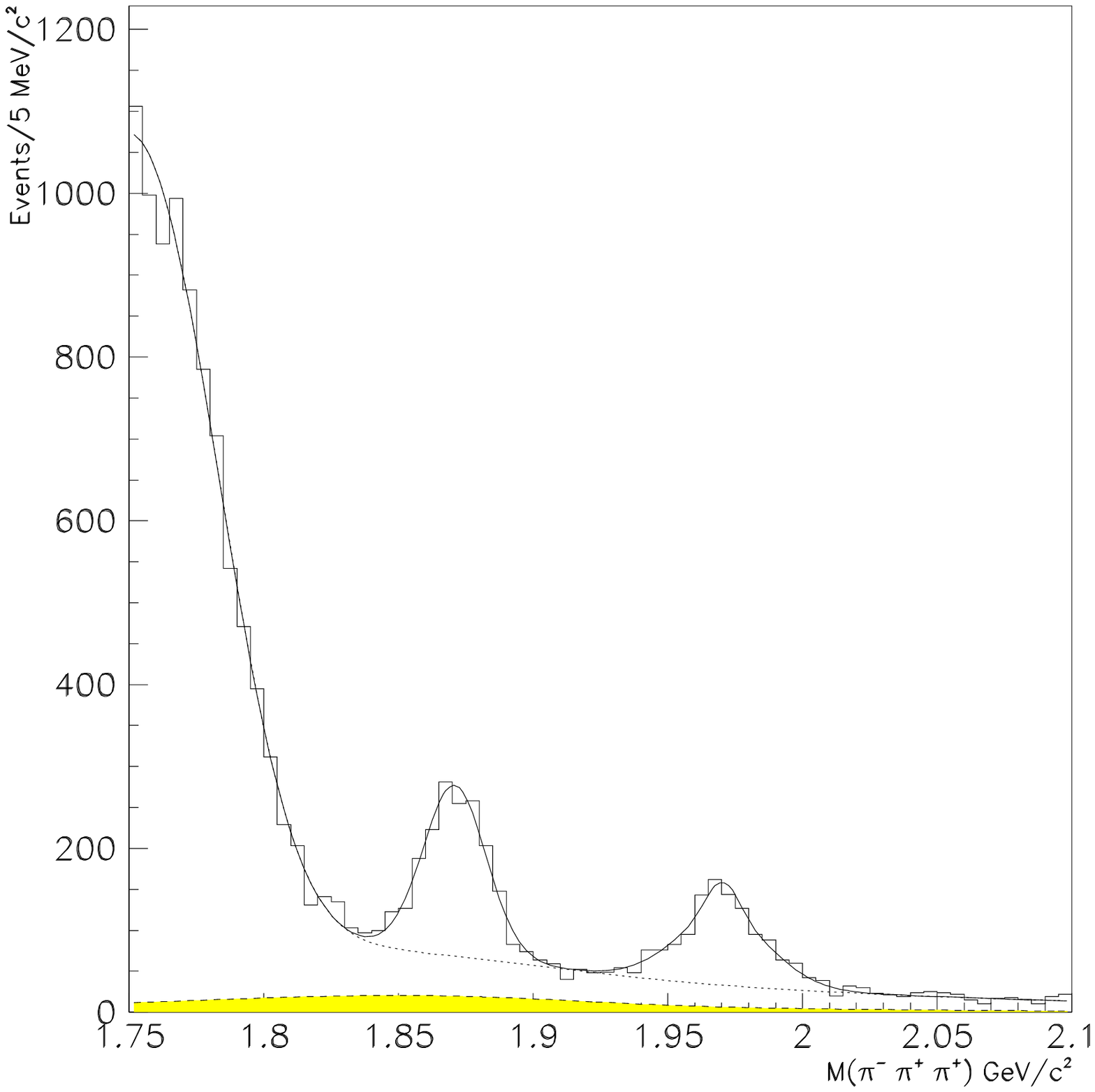,width=2.5in,height=1.5in,angle=0}
  \caption{{\bf Preliminary E791} data showing the three pion mass.
  Clear signals with $1240\pm 51~(\Dp)$
  and $858\pm 49~(\Dsp)$ events are seen after cuts to clearly require that
  the pions come from a vertex separated from the primary interaction point.
  Backgrounds from $\Dsp\to\etp\pi$ and $\Dz\to\Km\pip$ discussed in the text are in
  yellow.}
 \label{fig-e7913pi}
 \end{minipage}
}%  FIGURE
The sample was selected after making cuts aimed at identifying a clearly
separated $3\pi$ vertex.  No Cherenkov cut (relatively poorly simulated
in the Monte Carlo simulation control samples) was made to identify the pions.
Backgrounds from possible reflections from $\Dsp\to\etp\pip$ followed by
$\etp\to\rho\pi$ and from $\Dp\to\Km\pip$ (with additional $\pim$ from
the background) as well as from charmless, three pion combinations were
carefully estimated.

Branching ratios were normalised to $\Dp\to\Km\pip\pip$
(34,790$\pm$232 events) and to $\Dsp\to\phi\pip$ ($1038\pm 44$ events)
respectively
% (shown in figure \ref{fig-e791norms})
.  The
same cuts were applied to these normalisation samples as to the $3\pi$ data.
%
%\FIGURE{%
% \begin{minipage}{0.45\textwidth}
%  \epsfig{file=../E791/D/fig11_13.eps,width=2.9in,height=2.0in,angle=0}
%  \caption{E791 normalization signals:
%     $\Dp\to\Km\pip\pip (35400\pm 356$ events) 
%     $\Dsp\to\phi\pip 1038\pm 44$ events.)
%  }%  \caption
%  \label{fig-e791norms}
% \end{minipage}
%}%  FIGURE
%
Efficiencies were obtained from a full Monte Carlo simulation which included
appropriate distributions of production and resonant sub-channel structure in
the $D$ mesons.  The results
\begin{eqnarray*}
 BR\left({\Dp\rightarrow\pim\pip\pip       \over
            \Dp\rightarrow  \Km\pip\pip}   \right)
   &=&0.0329\pm 0.0015^{+0.0016}_{-0.0026} \\
 BR\left({\Dsp\rightarrow\pim\pip\pip      \over
            \Dsp\rightarrow\phi\pip}       \right)
   &=&0.247 \pm 0.028^{+0.019}_{-0.012}
\end{eqnarray*}
are compared in figure \ref{fig-e791br} with earlier experiments.  Though
agreement is generally good, a 3$\sigma$ discrepancy of the $\Dp$ rate with
respect to the E687 result is observed.
\FIGURE{%
 \begin{minipage}[t]{0.22\textwidth}
  \epsfig{file=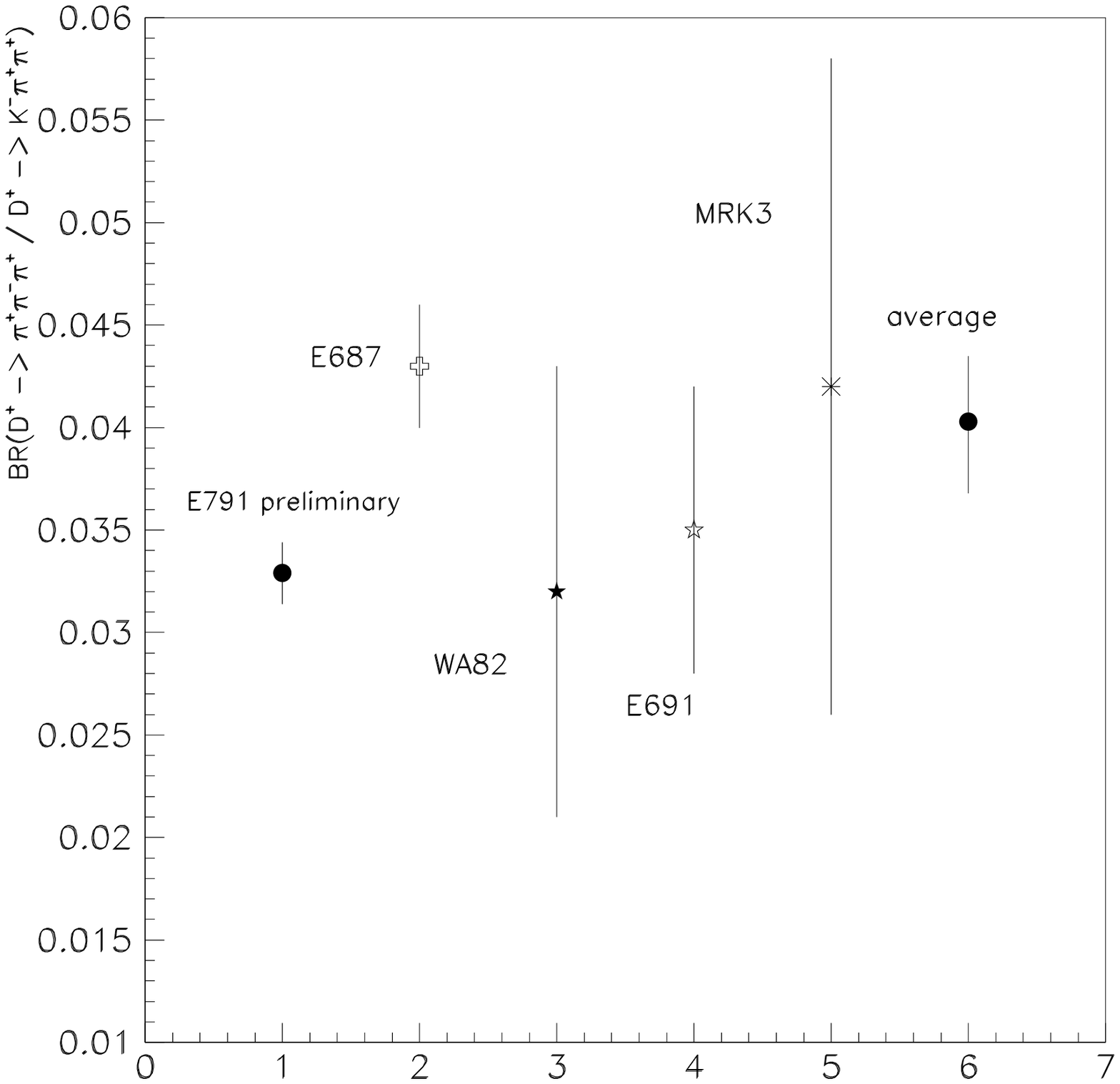,width=1.45in,height=1.25in,angle=0}
 \end{minipage}
 \begin{minipage}[t]{0.22\textwidth}
  \epsfig{file=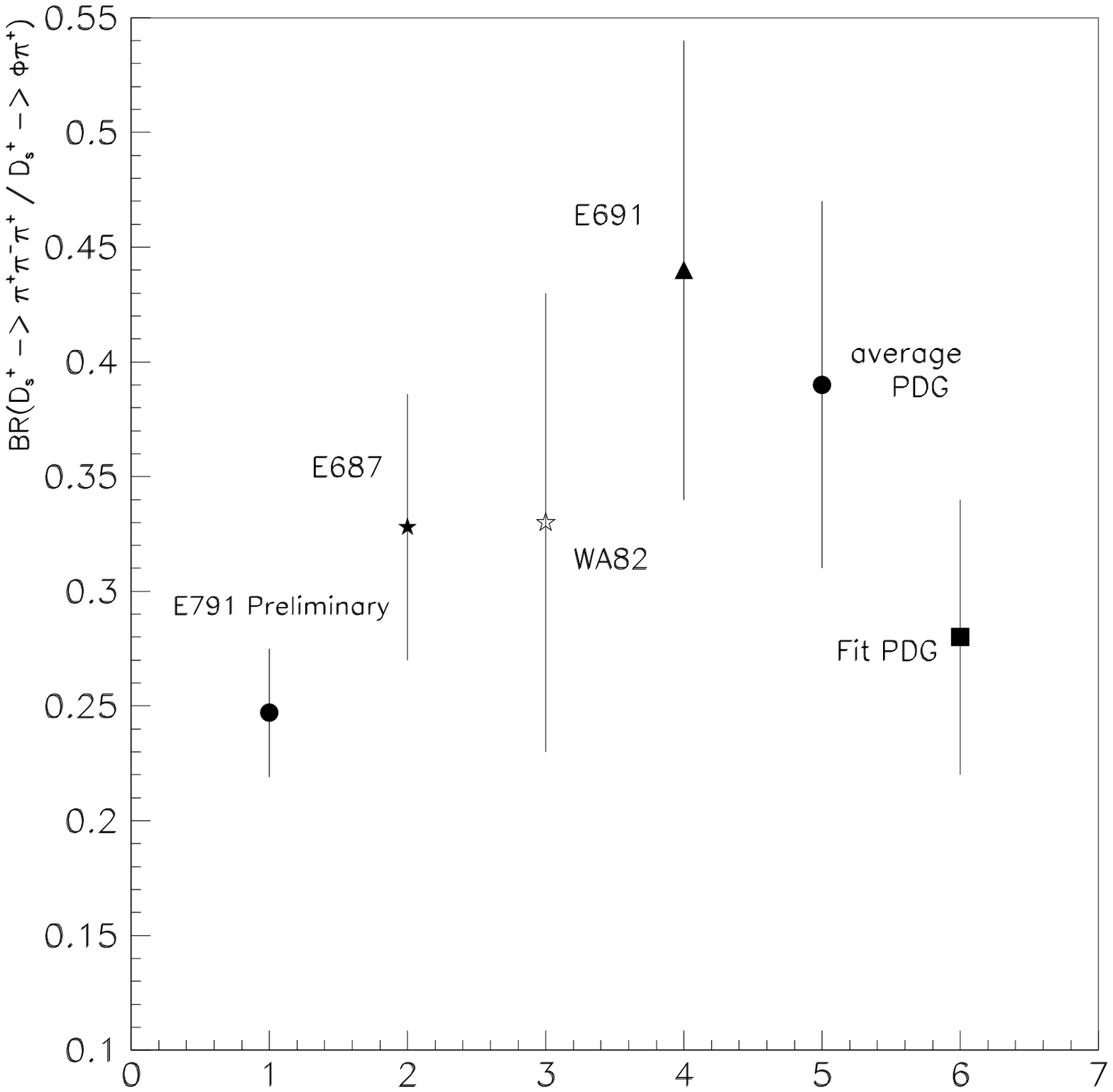,width=1.45in,height=1.25in,angle=0}
  \caption{Branching ratios for $\Dp$ and $\Dsp$ to three pions.}
  \label{fig-e791br}
 \end{minipage}
}%  FIGURE

\subsubsection{Resonant sub-channel analysis}

Figure \ref{fig-dalitz} shows the Dalitz plots for $\Dp$ and $\Dsp$ decays.
Each event with $\pim_1\pip_2\pip_3$ mass consistent with the appropriate $D$
meson appears with the squared invariant mass $s_{12}=M^2_{\pim_1\pip_2}$ plotted
against $s_{13}$.  The plots are folded as the two $\pip$'s are indistinguishable.
\FIGURE{%
 \begin{minipage}[t]{0.45\textwidth}
  \begin{minipage}[t]{0.45\textwidth}
   \epsfig{file=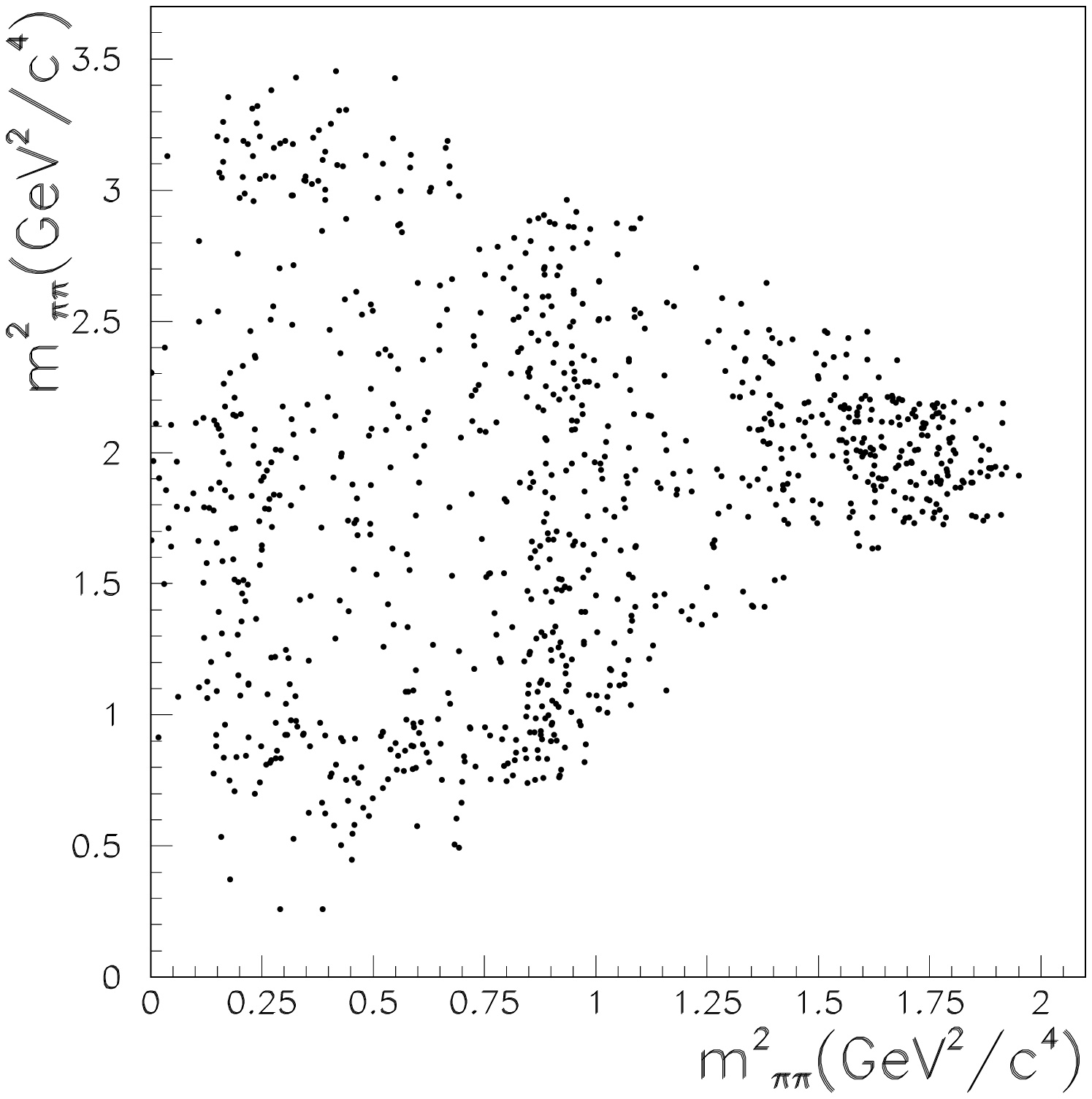,width=1.55in,angle=0}
  \end{minipage}
  \hspace{0.02\textwidth}
  \begin{minipage}[t]{0.45\textwidth}
   \epsfig{file=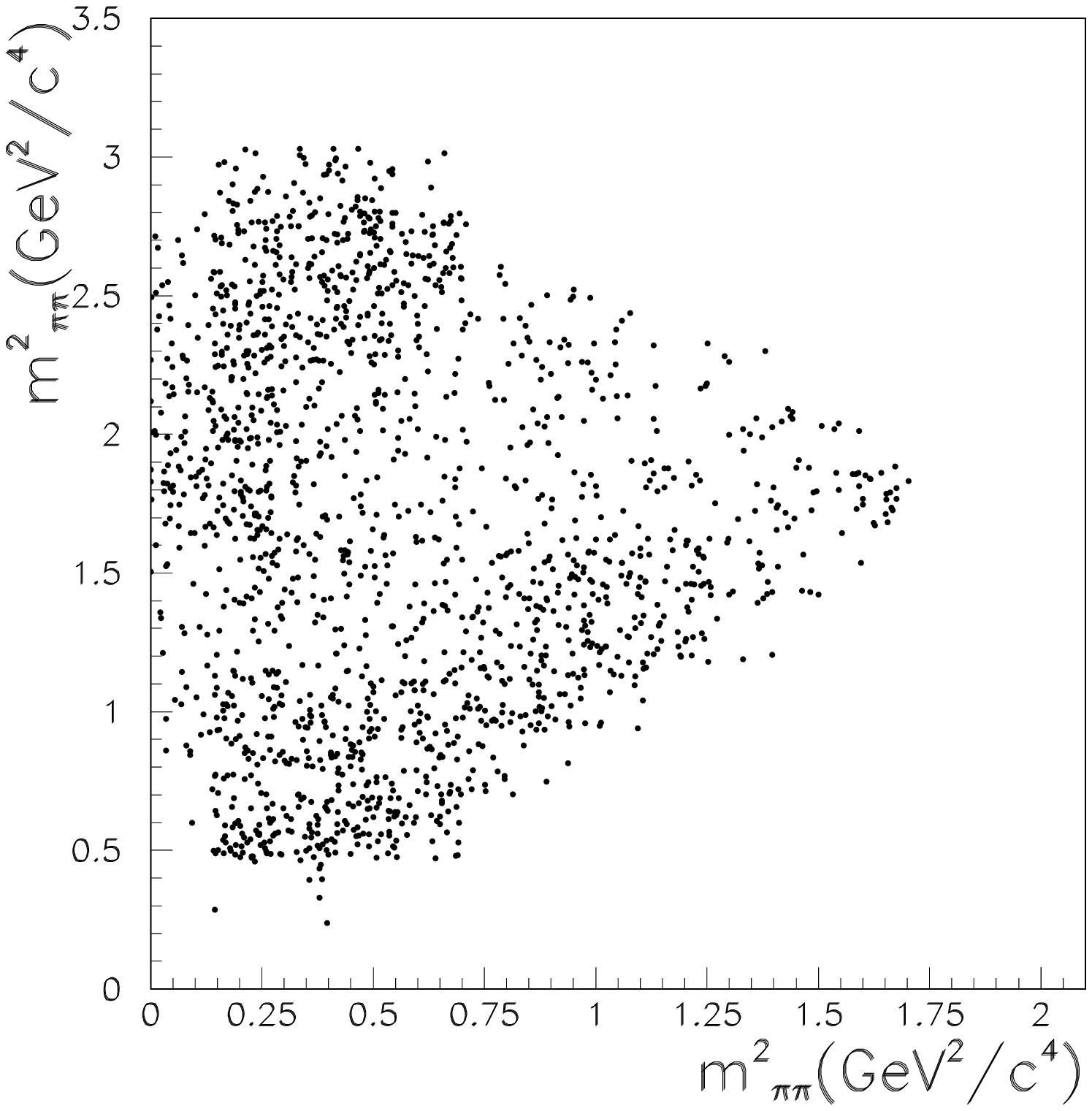,width=1.55in,angle=0}
  \end{minipage}
  \caption{Dalitz plots for $\Dsp$ (left) and $\Dp$ (right) decays to
  $\pim_1\pip_2\pip_3$.  Squared effective masses of the two $\pim\pip$
  combinations are plotted.  Each event is plotted twice.  In the text
  the abscissa is labelled $s_{12}$ and the ordinate $s_{13}$.}
  \label{fig-dalitz}
 \end{minipage}
}%  FIGURE
Prominent features evident in the $\Dsp$ Dalitz plot are a remarkable
concentration of events at the right apex of the plot (the {\sl ``Corcovado
Region"}) and clear bands corresponding to $\fz(980)$ decays to $\pim\pim$.
There appears to be little if any non resonant background.

The $\Dp$ Dalitz plot shows evidence for significant $\rhoz(770)\pi^{\pm}$
({\sl unlike the $\Dsp$ Dalitz plot}) and there is also a pronounced excess of
events along border, $s_{12}<0.3 (GeV/c)^2$.

As in previous analyses of these decays, these plots were fitted with a
{\sl coherent} linear sum of amplitudes $\Aj$ with complex coefficients
$\aj e^{i\deltj}$ - one for each sub-channel process $j$.  Each
$\Aj$ (Bose symmetrised wrt $2\leftrightarrow 3$)
consisted of an appropriate angular dependent form for the $\pim\pip$ partial
wave with a Breit-Wigner (BW) amplitude for each of the resonances evident in
the data
\footnote{The $\fz(980)$ state, since its nominal mass is near the $KK$
threshold, was also represented by a coupled channel form whose results are not
reported here.}.
Non resonant decay to $3\pi$ was taken to be $s-$wave with amplitude
independent of position in the Dalitz plot.  Distributions for the various
backgrounds and efficiency as function of position on the plot were estimated
from Monte Carlo studies.

Fits were made to determine magnitude $\aj$ and relative phase $\deltj$ for
each process.  In the case of scalar mesons with poorly determined 
mass and width these quantities were also allowed to vary in the fits.
``Resonant fractions" $\fj$ were derived from these parameters:
 \[
   \fj={\int ds_{12}ds_{13}\left|\aj e^{i\deltj}\Aj\right|^2 \over
        \int ds_{12}ds_{13}\left|\sum_j \aj e^{i\deltj}\Aj\right|^2}
 \]
It is noted that interference means that the sum of $\fj$'s is not necessarily
unity as the integrals do not extend to $\infty$, however it should be
approximately so.  Interference also means that the significance of a
resonant fraction $f$ is not necessarily the same as that of an
amplitude $a$.

\subsubsection{$\Dsp$ sub-channels results.}

For the $\Dsp$ an excellent fit was found with decay channels indicated in
table \ref{tab-dsp}.  The projection onto the $\pim\pip$ mass for this fit is
shown in figure \ref{fig-dsp2pi}.

\TABLE[ht]{%
 \begin{minipage}{\textwidth}
 \centering
 \begin{tabular}{ c c c c }
 \bf Mode       & \bf Amplitude $\aj$ & \bf Phase $\deltj$ (radians)
 & \bf Fraction $\fj$
 \\ 
   $\fz(980)\pi^+$& 1(fixed)&0(fixed) &0.57 $\pm$0.04 $\pm$0.05
 \\
  $     NR     $ &0.07 $\pm$0.10 $\pm$0.03 &2.00 $\pm$1.33 $\pm$0.59
&0.01 $\pm$0.01 $\pm$0.01
 \\
  $\rho^0(770)\pi^+$  &0.33 $\pm$0.06 $\pm$0.16 &1.62 $\pm$0.44$\pm$0.09
&0.06
  $\pm$0.02 $\pm$0.04
 \\
   $\f2(1270)\pi^+$  &0.58 $\pm$0.06 $\pm$0.01 &2.10 $\pm$0.20
$\pm$0.45&0.19 $\pm$0.03 $\pm$0.01
 \\
   $\fz(1370)\pi^+$  &0.74 $\pm$0.09 $\pm$0.02 &3.29 $\pm$0.19 $\pm$0.45
&0.31 $\pm$0.06  $\pm$0.02
 \\
  $\rho^0(1450)\pi^+$ &0.28 $\pm$0.08 $\pm$0.01 &2.40 $\pm$0.38 $\pm$0.35
&0.04 $\pm$0.02  $\pm$0.01
 \\
 \end{tabular}
 \caption{{\bf Preliminary} results of sub-channels fit to decays of $\Dsp$ from E791.}
 \label{tab-dsp}
 \end{minipage}
}%  \TABLE
\FIGURE[ht]{%
 \begin{minipage}{0.45\textwidth}
 \epsfysize=1.0in
%\centerline{\epsffile{../E791/D/fig29.eps}}
 \centerline{\epsffile{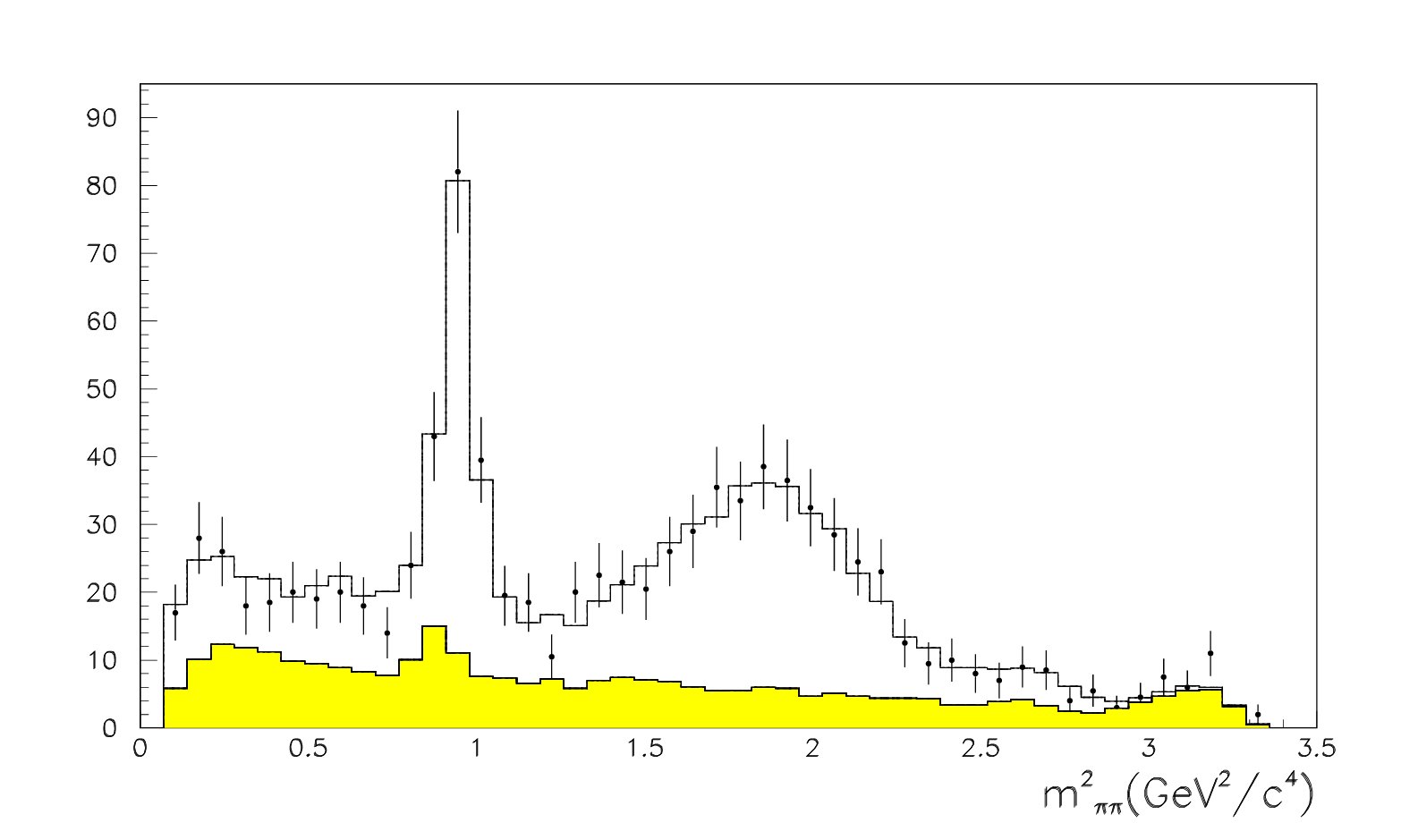}}
 \caption{Projection of fit to E791 $\Dsp$ decays to three pions detailed in
          table \ref{tab-dsp}.  This result is {\bf preliminary}.}
 \label{fig-dsp2pi}
 \end{minipage}
}%  \FIGURE

The $\fz(980)\pi^{\pm}$ mode is the dominant one and the signal in the
plot is striking.  The fraction of
non resonant decay is close to zero but on the other hand, both
$\rhoz(770)\pi^{\pm}$ and $\rhoz(1450)\pi^{\pm}$ have significant amplitudes
(but less significant fractions.)  This is the first evidence for $\rho\pi$
decay of $\Dsp$ and could indicate either contribution from the annihilation
diagram or from inelastic FSI.  Despite very significant interference, the
sum of fractions is about 1.2 - close to unity.

\subsubsection{$\Dp$ sub-channels results.}

A similar fit to the $\Dp$ Dalitz plot was made.
The fit is shown in figure \ref{fig-dp2pi} (to the left).  This fit is
poor in the low mass $\pip\pim$ region, and has a number of other
unsatisfactory features.
\FIGURE[ht]{%
 \begin{minipage}[t]{0.45\textwidth}
  \begin{minipage}[t]{0.45\textwidth}
   \epsfxsize=1.55in
%  \centerline{\epsffile{../E791/D/fig36.eps}}
   \centerline{\epsffile{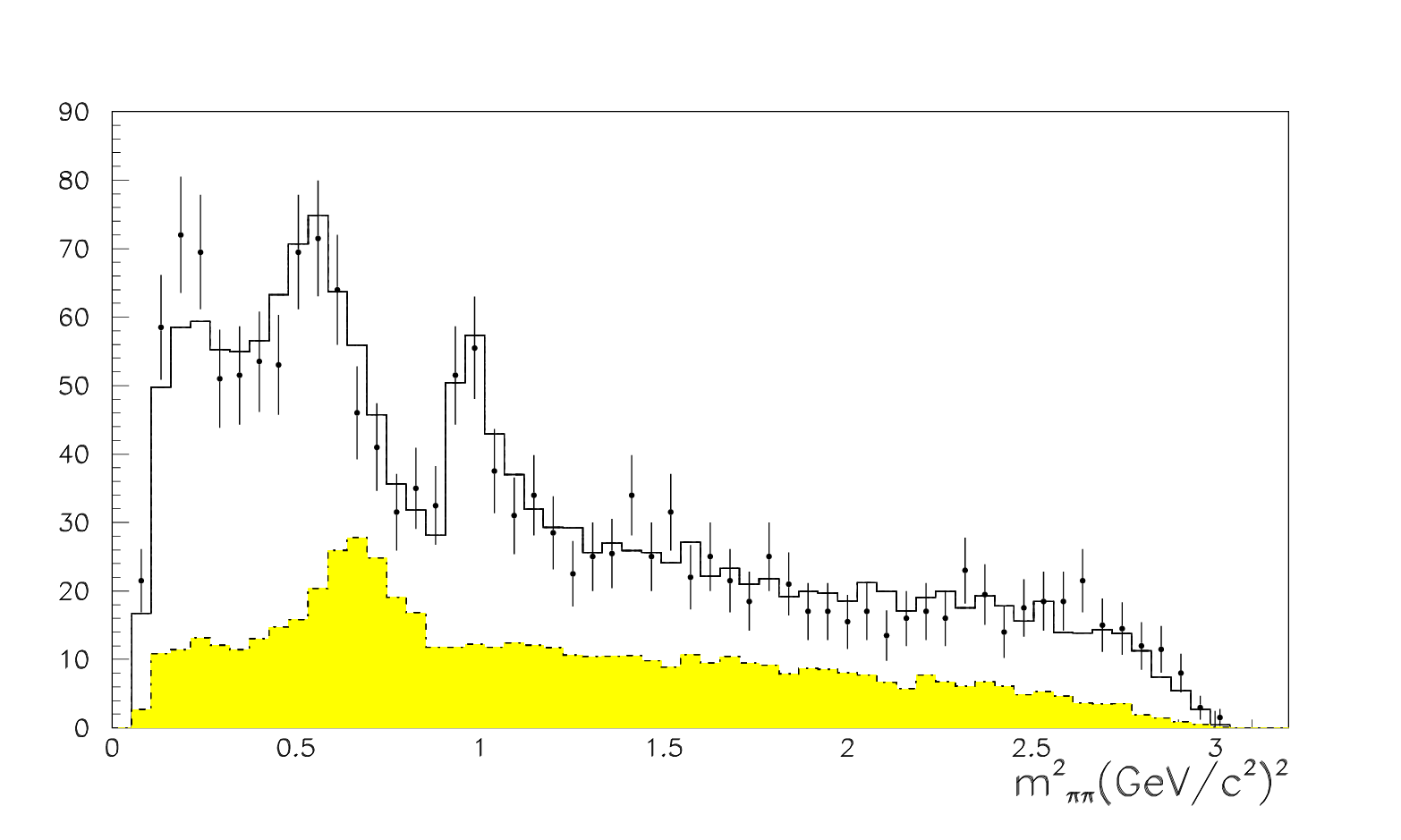}}
  \end{minipage}
  \hspace{0.03\textwidth}
  \begin{minipage}[t]{0.45\textwidth}
   \epsfxsize=1.55in
%  \centerline{\epsffile{../E791/D/fig38.eps}}
   \centerline{\epsffile{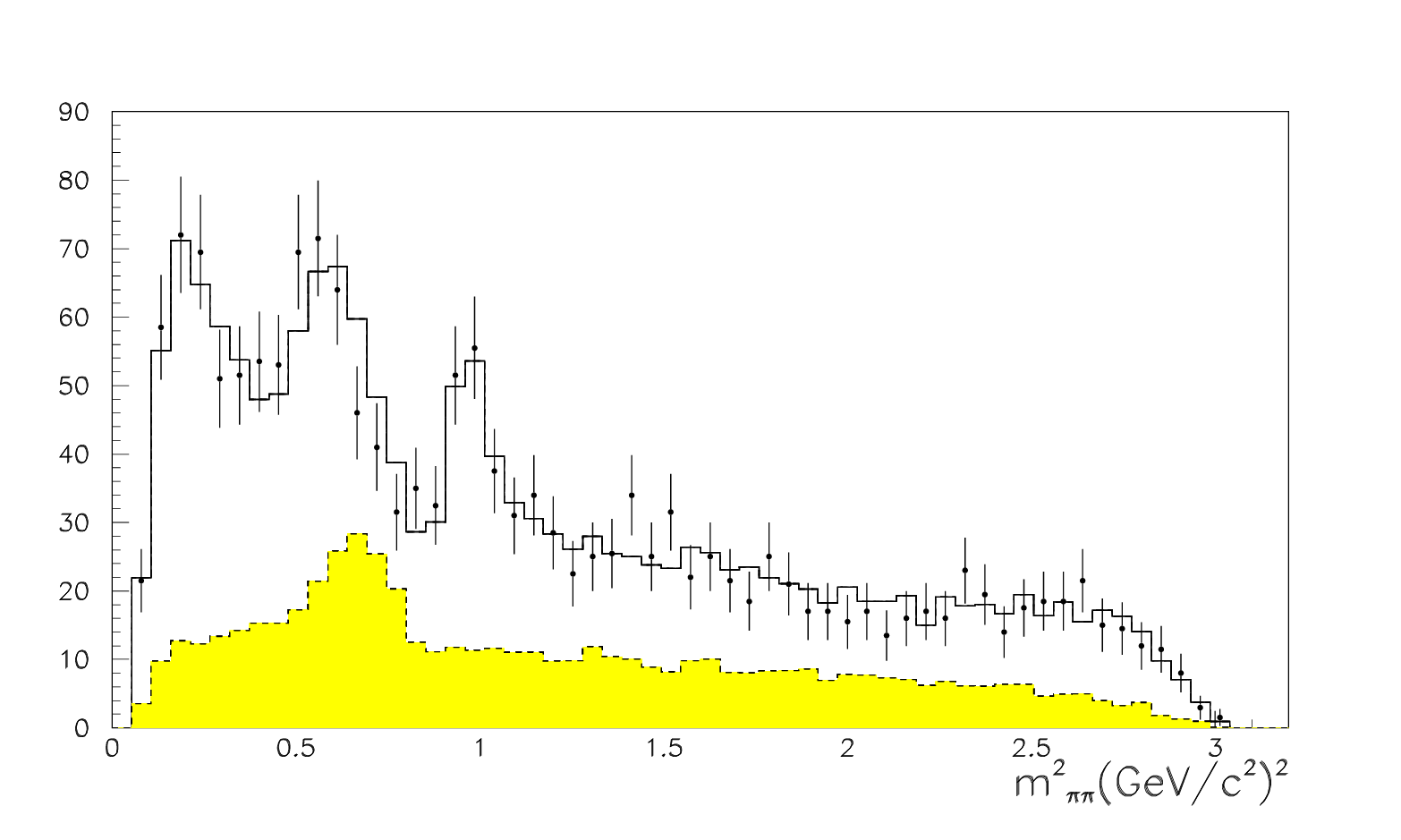}}
  \end{minipage}
  \caption{Fit (solid line) to $3\pi$ Dalitz Plot for $\Dp$ data (crosses)
           projected onto $\pim\pip$ mass.  On the left, no additional 
           $s-$wave amplitude was added to the known resonances included
           in table \ref{tab-dp}.  The fit on the right included an $s-$wave
           BW amplitude with mass and width completely free to vary.  The fit
           converged on a mass of $486^{+28}_{-26}$ MeV and width of
           $351^{+50}_{-43}$ MeV.  Estimated backgrounds
           from other $D$ decays are shown in yellow.  This result is
           {\bf preliminary}.}
  \label{fig-dp2pi}
 \end{minipage}
}%  \FIGURE
First, decay to $\rhoz(1450)\pip$ was more significant than
$\rhoz(770)\pip$ - a somewhat surprising situation. Also,
the non resonant $3\pi$ decay was the dominant process, somewhat
different from the $\Dsp$ decay.

To achieve a better fit, an additional $s-$wave amplitude was added.  A
BW form was chosen with mass and wdth completely free to vary.  The fit
converged with mass $486^{+28}_{-26}$ MeV and width $351^{+50}_{-43}$ 
MeV.  This amplitude was named as ``$\sigma$".
In the fit with $\sigma$ the $\sigma\pip$ contribution
dominated, non resonant decay was very small and the $\rhoz(1450)\pip$
fraction was much less than $\rhoz(770)\pip$.  Also the phase difference
between $\fz(980)$ and $\f2(1270)$ was about {\bf two radians} - much the
same as for the $\Dsp\to\pim\pip\pip$ decay.  For the fit without $\sigma$
this was only about one radian.  This fit appears to be preferred to 
that without $\sigma$.  It even restores some symmetry with respect to
the $\Dsp$ decay in that quasi two body modes again dominate.
Preliminary results from the fit with this state are given in table \ref{tab-dp}.
\TABLE[ht]{%
 \begin{minipage}{\textwidth}
  \centering
  \begin{tabular}{ c c c c }
  \bf Mode       & \bf Amplitude $\aj$ & \bf Phase $\deltj$ (radians)
  & \bf Fraction $\fj$
  \\
   $\sigma\pi^+$ &1.27 $\pm$0.13 $\pm$0.22 &3.57 $\pm$0.12
   $\pm$0.17 &0.49 $\pm$0.09 $\pm$0.13
  \\
    $\rho^0(770)\pi^+$& 1(fixed)&0(fixed) &0.30 $\pm$0.03 $\pm$0.03
  \\
   $     NR     $ &0.50 $\pm$0.13 $\pm$0.12 &1.23 $\pm$0.26
   $\pm$1.41 &0.08 $\pm$0.04 $\pm$0.04
  \\
   $\fz(980)\pi^+$  &0.50 $\pm$0.06 $\pm$0.08 &2.86 $\pm$0.18
   $\pm$0.34 &0.08 $\pm$0.02 $\pm$0.02
  \\
   $\f2(1270)\pi^+$  &0.80 $\pm$0.07 $\pm$0.09 &0.91 $\pm$0.13
   $\pm$0.24 &0.19 $\pm$0.02 $\pm$0.03
  \\
   $\fz(1370)\pi^+$  &0.30 $\pm$0.09 $\pm$0.09 &1.92 $\pm$0.27
   $\pm$0.11 &0.03 $\pm$0.02 $\pm$0.02
  \\
   $\rho^0(1450)\pi^+$ &0.19 $\pm$0.08 $\pm$0.12 &5.86 $\pm$0.50
   $\pm$1.64 &0.01 $\pm$0.01 $\pm$0.02
  \end{tabular}
  \caption{{\bf Preliminary} results of sub-channels fit to decays of $\Dp$
           from E791.  Fit includes a low mass $\pim\pip$ state ``$\sigma$".}
  \label{tab-dp}
 \end{minipage}
}%  \TABLE

\subsubsection{Results pertinent to light quark states.}
The $\sigma$ state observed by E791 in $\Dp$ decay could be the
$\fz(400-1200)$ state in the particle data group tables.  In addition to
the $s-$wave BW form, E791 fit this as a $p-$wave as well as a
$d-$wave resonance, in each case with a significantly worse likelihood.  
Furthermore, to determine if the state exhibited significant phase variation
as a true resonance should, they made a fit with a BW mass envelope (a bump)
but without the BW $\pim\pip$ mass dependent phase variation.  This fit also
gave a worse likelihood and an unphysically large sum of fractions.

It should be pointed out that though E791 sees a significant improvement
in their fit to $\Dp$ decay with the $\sigma$, they did not attempt to 
fit their data with other possible non resonant models such as an
effective range parametrization or amplitudes measured in earlier
$\pi\pi$ scattering experiments.  Their evidence for a $\sigma$ state
therefore, though very interesting, has to be considered with this 
caveat in mind I believe.

The strong signals for both $\fz(980)$ and $\fz(1370)$ in $\Dsp$ decay
are the clearest manifestations seen to date.  In each case, E791
extracted information which adds to that published \cite{pdg:1999}.  
Table \ref{tab-scalars} summarises preliminary parameter values found 
for these states as well as for the ``$\sigma$" when treated as 
$s-$wave BW's. \footnote{This parametrisation is acknowledged to be
unsuitable in general for broad overlapping resonances with a number
of open channels for decay, such as the $\sigma$ might represent.
However E791 presents these parameters for empirical reasons.}
\TABLE[ht]{%
 \begin{minipage}{0.45\textwidth}
  \begin{tabular}{ c c c }
  \bf Parameter   & E791 Value        & \bf PDG Value  \\ \hline
  \multicolumn{3}{l}{\bf From $\Dsp\to \fz(980)\pip$:}     \\ 
      Mass (MeV)  & $978 \pm 4$       &  980 $\pm$ 10  \\
      Width (MeV) & $44 \pm 5$        &   40 to 100    \\ \\
  \multicolumn{3}{l}{\bf From $\Dsp\to \fz(1370)\pip$:}    \\
      Mass (MeV)  & $1440 \pm 19$     & 1200 to 1500   \\
      Width (MeV) & $165 \pm 29$      &  200 to 500    \\ \\
  \multicolumn{3}{l}{\bf From $\Dp\to ``\sigma"\pip$:}     \\ 
      Mass (MeV)  & $486^{+28}_{-26}$ & 400 to 1200    \\
      Width (MeV) & $351^{+50}_{-43}$ & 600 to 1000
  \end{tabular}
 \end{minipage}
 \caption{Parameters from E791 Dalitz plot fits relevant to light quark
          scalar mesons.  Masses and widths reported here relate to
          $s-$wave BW forms.  Results are {\bf preliminary}.}
 \label{tab-scalars}
}%  \TABLE

\def\rhoz{\rho^{\circ}}
\def\Aj{{\cal A_{\rm j}}}
\def\llik{{\cal L}}
\def\Ps{{\cal P_{\rm s}}}
\def\PB{{\cal P_B}}
\def\aj{\Red{a_{\rm j}}}
\def\deltj{\Red{\delta_j}}
\def\fj{\Red{f_{\rm j}}}
\def\bi{\Red{b_{\rm i}}}
%___________________________________________________________________
\section{\bf Multidimensional Analysis of $\Lc\to p\Km\pip$ Decays}

The first amplitude analysis of the decay of a charmed baryon has
been reported by the E791 collaboration \cite{E791-lamcpol:1999}.  A
sample of almost 900 events was used to determine not only the
relative strengths and phases of resonances in the final state,
but also the $\Lc$ production polarisation (as a function of $p_T$).

Understanding the resonant decomposition of the final state helps
unravel the relative importance of the lowest order processes
in figure~\ref{fig-baryon_decay}.
\FIGURE{%
\begin{minipage}{0.40\textwidth}
  \centering
  \epsfig{file=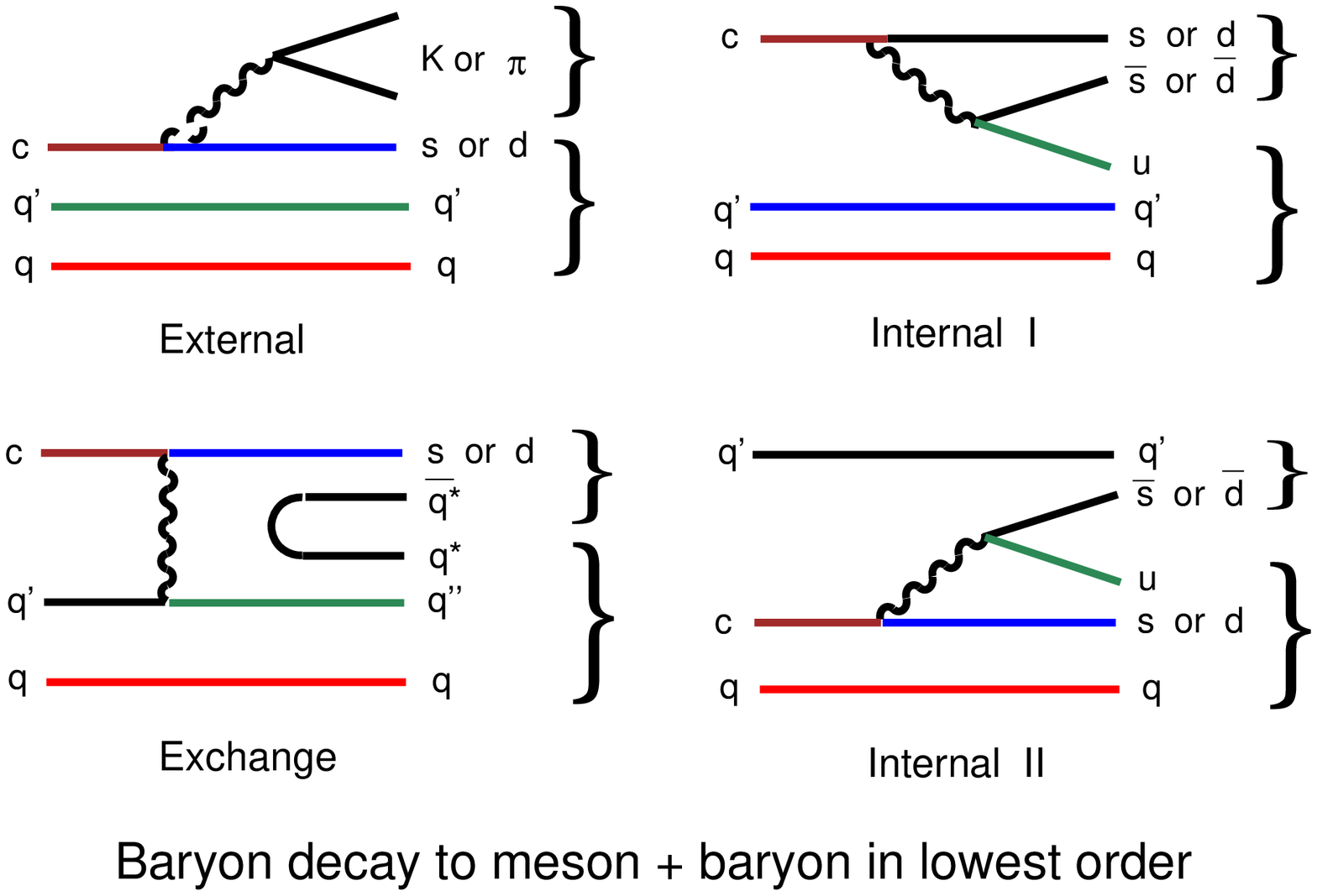,width=2.8in,angle=0}
  \caption{Leading tree level processes contributing to charmed baryon
           decays to baryon + meson final states.}
\end{minipage}
\label{fig-baryon_decay}
}% \FIGURE
The ``exchange" diagram can contribute to $p\Kst$, $\Lambda^{\ast}\pi$,
$\Sigma^{\ast}\pi$ or $pK\pi$ modes.  However, for the $\Delta^{++}\Km$
mode it is the only diagram possible.  In meson decays, the exchange
mechanism is expected (and found) to be relatively weak due to helicity
suppression which is not a factor in baryon decays.

\subsection{\bf E791 Sample}

The sample used by E791 in this analysis is shown in figure~\ref{fig-e791lamc}.
The signal had $886\pm 43$ events ($\sim$ 20 standard deviations) and 
the total number of background events in the plot was $1384\pm 49$.
\FIGURE{%
\begin{minipage}{0.4\textwidth}
  \centering
  \epsfig{file=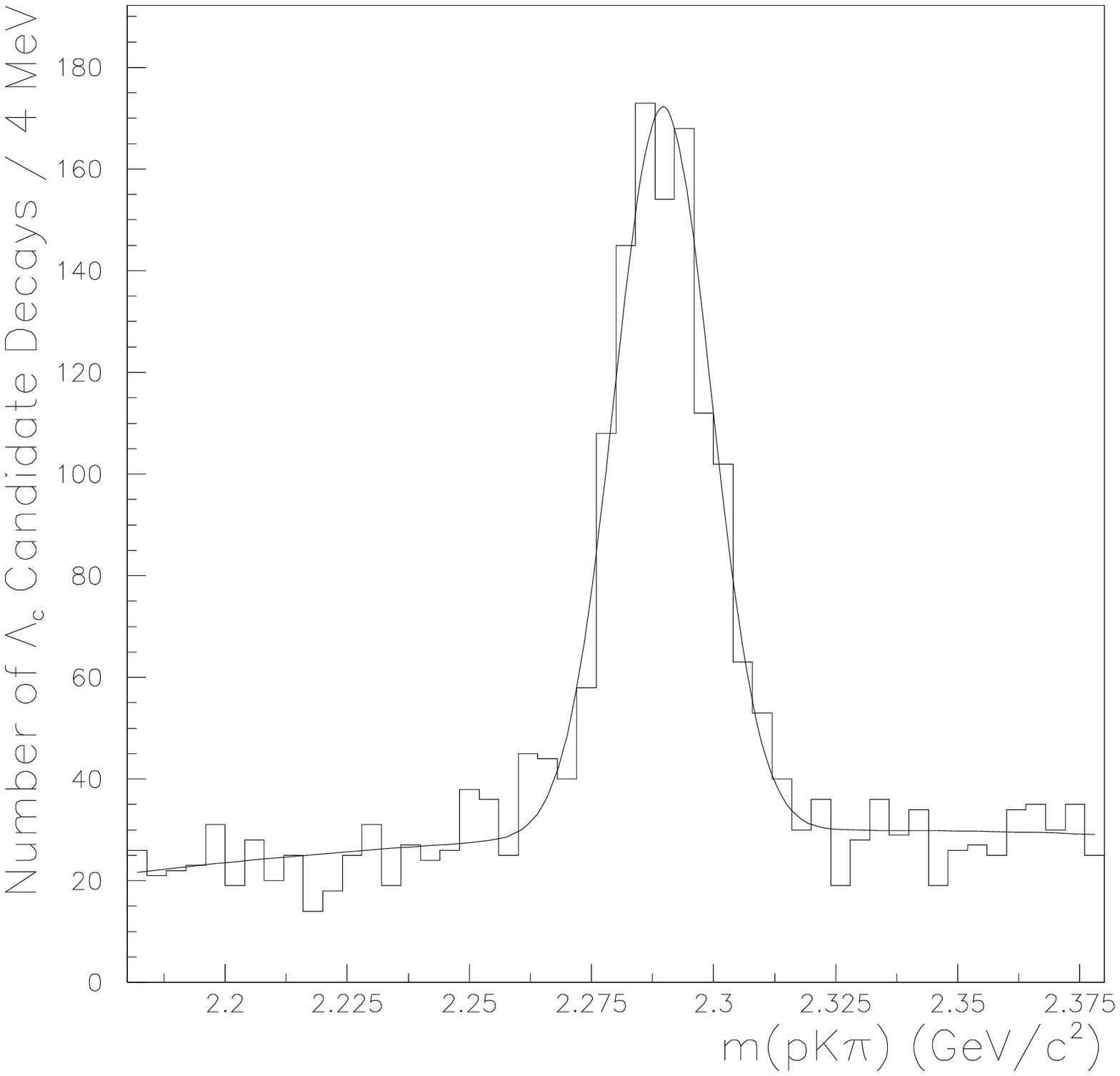,width=2.5in,height=1.2in,angle=0.0}
  \caption{E791 sample used in the amplitude fit.}
  \label{fig-e791lamc}
\end{minipage}
}% \FIGURE
\subsection{\bf Basic Fit Formalism}

This analysis was more complex than that for $\Dp (\Dsp)\to\pim\pip\pip$
though a comparison is helpful.
Both parent and daughter baryons have spin-parity $J^P=\half^+$ making
initial polarisation and helicity states important.
Also, five rather than two dimensions were required (two Dalitz plot 
variables as well as the orientation of the decay plane.)

The density function used in the $\Dp$ and $\Dsp$ analyses to describe
the Dalitz plot distribution $\Ps$ was a sum over sub channels of
amplitudes $R_j=\aj~e^{i\deltj}\Aj(M_j)$, each with a complex coefficient
$\aj~e^{i\deltj}$.  In the $\Lc$ analysis, each sub channel amplitude
had to be replaced by a sum over helicities:
\[
\begin{array}{l l}
 R_j&=~~~\left({1+P_{\Lc}\over 2}\right)\times \\
    &    \left(%
         \left|\sum_j\Aj(M_j)\alpha^j_{\half, \half}\right|^2
     +   \left|\sum_j\Aj(M_j)\alpha^j_{\half,-\half}\right|^2 \right) \\
    &+~~~\left({1-P_{\Lc}\over 2}\right)\times \\
    &    \left(
         \left|\sum_j\Aj(M_j)\alpha^j_{\half, \half}\right|^2
     +   \left|\sum_j\Aj(M_j)\alpha^j_{\half,-\half}\right|^2 \right)
\end{array}
\]
with coefficients $\alpha^j_{m,\lambda_p}$, labelled by $m$
(spin projection of $\Lc$ on the polarisation axis $\vec P_{\Lc}$)
and $\lambda_p$ (helicity of proton in $\Lc$ rest frame.)

The background density and efficiency were determined empirically in 5
dimensions from the observed density of data in side bands of the
$p\Km\pip$ mass plot using the ``nearest next neighbours" technique.

A fit to $p\Kstbar(890)$, $\Delta^{++}(1232)\Km$,
$\Lambda(1520)\pip$ \& $NR$ modes required 34 parameters including
amplitudes and phases for each helicity state of each sub channel,
polarisations in each of three $p_T$ ranges.  Also included in the fit
were parameters describing the assumed signal to background ratio
and shape in figure~\ref{fig-e791lamc}.

\subsection{Results of the fit}

Figure \ref{fig-e791L4} shows the fit projected onto the two body mass
projections and the three angles which define the decay orientation.
Also shown are the measured polarisations in each $p_T$ range.
The fit describes the data quite well except for the low mass
$\Km p$ region which shows an excess in the data above the fit.
Various attempts to parametrise this with low mass $\Lambda$ and
$\Sigma$ states were unsuccessful as each new state required an
unacceptable increase in the number of parameters.  The polarisation
shows a trend to larger negative values with increasing $p_T$.

\FIGURE[ht]{%
 \hbox{%
  \epsfig{file=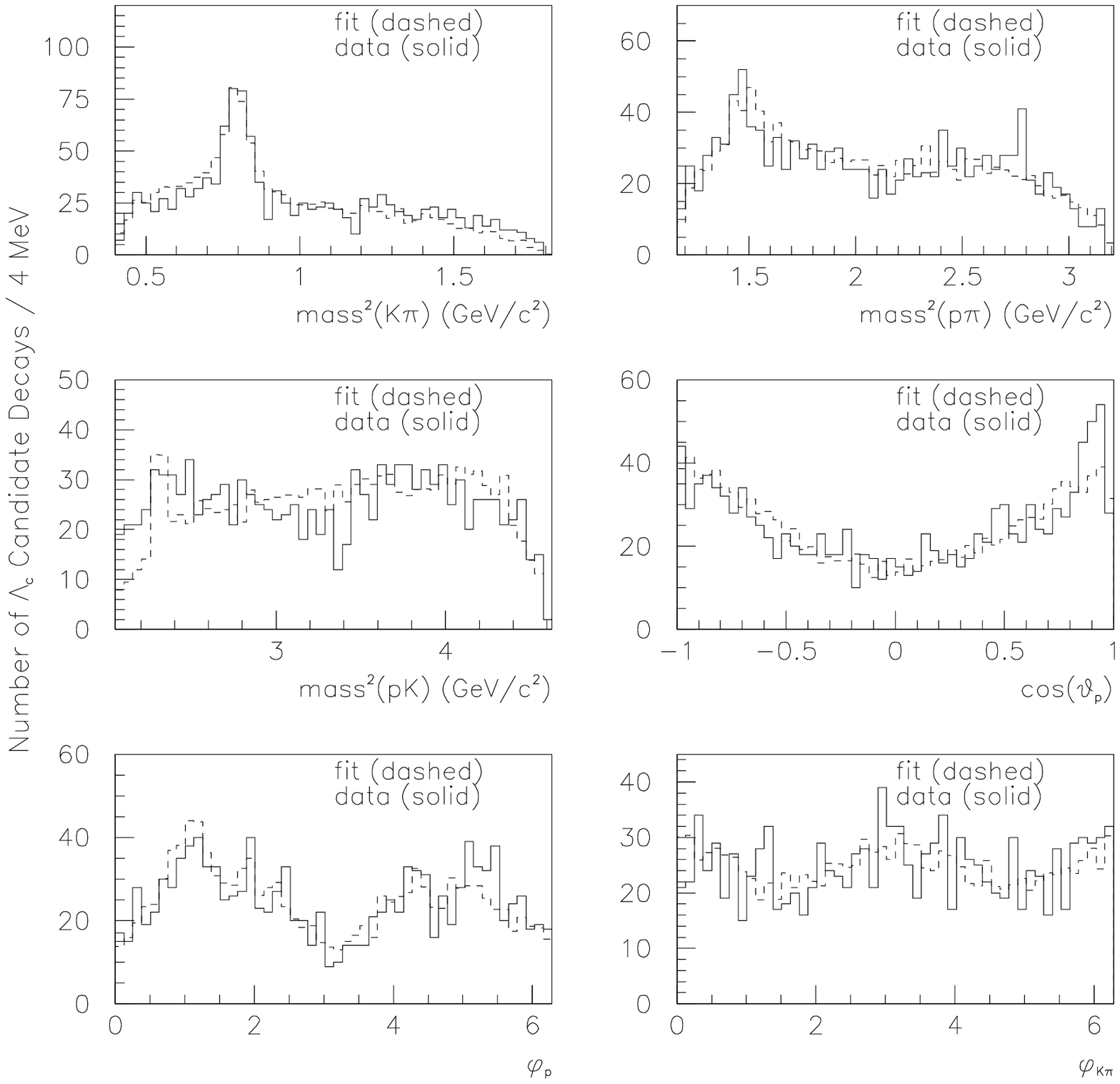,width=1.70in,angle=0.0}
  \hspace{0.01\textwidth}
%
% Now the right hand box:
% \epsfig{file=../E791/L/fig3.eps,width=0.8in,height=1.65in,angle=0.0}
  \epsfig{file=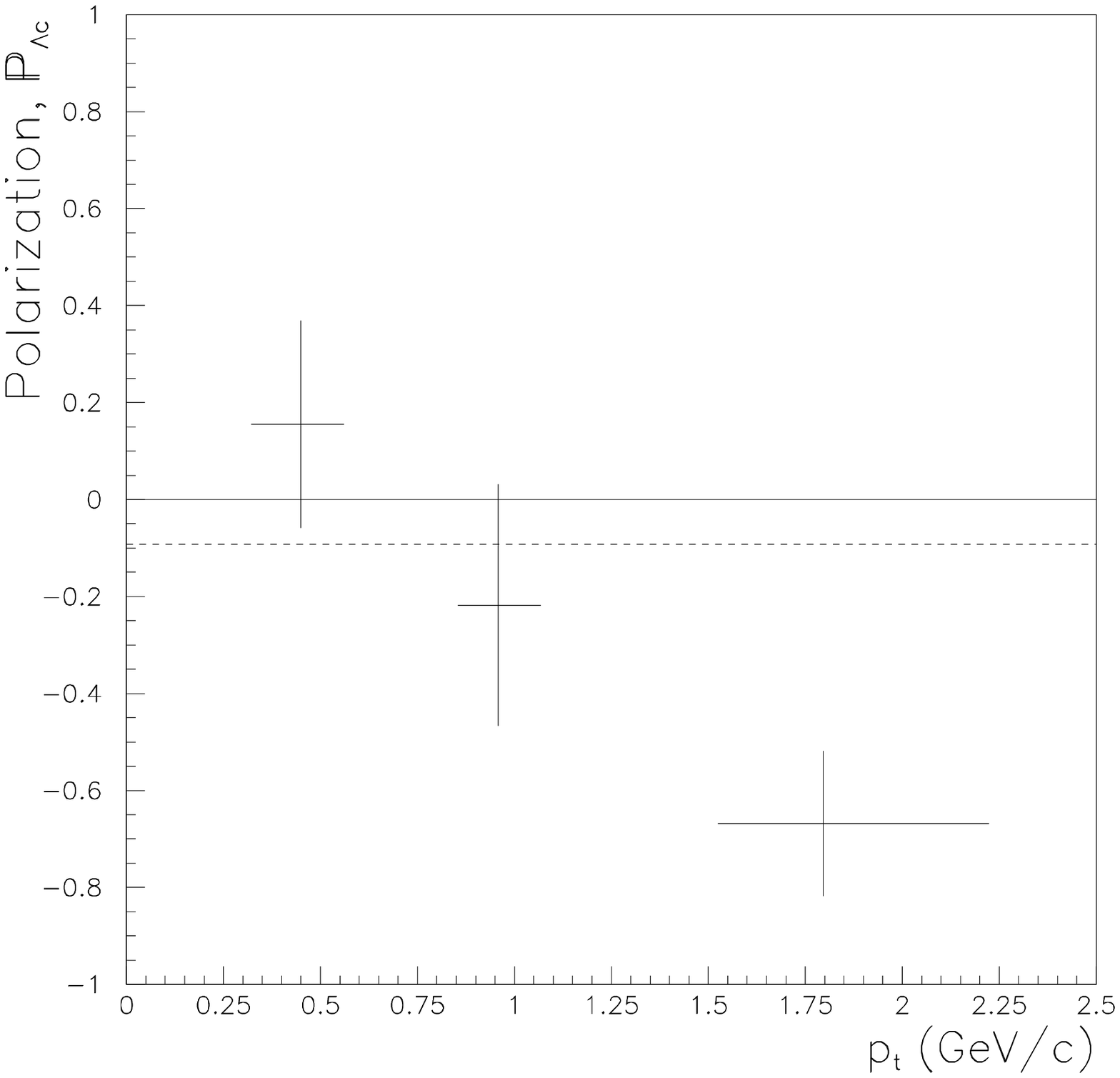,width=0.8in,height=1.65in,angle=0.0}
 }%  hbox
  \caption{Plot on the left shows projections of two body masses
           and angles.  Solid line histograms are data in the
           range $2265<M(p\Km\pip)<2315$~MeV/c$^2$.  Dashed lines
           represent the fit in the same range.  Plot at right
           shows polarisation $P_{\Lc}$ vs. $p_T$ from fit.}
  \label{fig-e791L4}
}% \FIGURE

Resonant sub channel fractions were computed as in the $\Dp(\Dsp)$
decay analysis and are summarised in table \ref{tab-e791L1}.

\TABLE[ht]{%
 \centering
 \begin{minipage}{0.4\textwidth}
 \begin{tabular}{ c l }
 \bf Mode & (\% of $p\Km\pip$)
 \\ \hline
 $p\Kstz(890)$
 & $19.5\pm 2.6\pm 1.8$
 \\
 $\Delta^{++}(1232)\Km$
 & $18.0\pm 2.9\pm 2.9$
 \\
 $\Lambda(1520)\pip$
 & $ 7.7\pm 1.8\pm 1.1$
 \\
 $NR$
 & $54.8\pm 5.5\pm 3.5$
 \\ \hline
 \end{tabular}
 \caption{Resonant fractions from $\Lc$ decay fit.  Sum is
          close to unity.}
 \label{tab-e791L1}
 \end{minipage}
}% \TABLE
The sum of fractions in the table is close to unity.  The non
resonant three body decay channel is strikingly large and dominant.
The $\Delta^{++}\Km$ subchannel is as large as $p\Kst$ and is
significant indicating that $W-$exchange is indeed an important
contribution to the decay.

\subsection{\bf Branching Ratios wrt $p\Km\pip$}

Allowing for unseen decay modes of the resonances involved, these
fractions can be converted into branching ratios with respect to
the $p\Km\pip$ mode.  They are compared in table \ref{tab-e791L2}
with previous measurements \cite{pdg:1999} and agreement seems to be
quite good.

Earlier measurements were made with much smaller samples and
resulted from fits to mass projections, ignoring interference
effects.  When correlations between channels are properly
taken into account, as in the E791 analysis, the uncertainties
grow.  Therefore, errors reported by E791 appear to be comparable
to those of NA32 for example \cite{Bozek:1993kq} even though a
much larger sample and more sophisticated analysis was used to
obtain these results.

\TABLE{%
\begin{minipage}[ht]{\textwidth}
 \centering
 \begin{tabular}{ c c c c }
 \bf Mode & \bf E791 & \bf NA32 & \bf ISR (SFM)
 \\ \hline
 $p\Kst(890)$
 & $0.29\pm 0.04\pm 0.03$
 & $0.35^{+0.06}_{-0.07}\pm 0.03$
 & $0.42\pm 0.24$
 \\
 $\Delta^{++}(1232)\Km$
 & $0.18\pm 0.03\pm 0.03$
 & $0.12^{+0.04}_{-0.05}\pm 0.05$
 & $0.40\pm 0.17$
 \\
 $\Lambda(1520)\pip$
 & $0.15\pm 0.04\pm 0.02$
 & $0.09^{+0.04}_{-0.03}\pm 0.02$
 &
 \\
 $NR$
 & $0.55\pm 0.06\pm 0.04$
 & $0.56^{+0.07}_{-0.09}\pm 0.05$
 &
 \\ \hline
 \end{tabular}
 \caption{Comparison of branching ratios for $\Lc$ decays to $p\Km\pip$
       subchannels.  First error is statistical, second is systematic.}
 \label{tab-e791L2}
\end{minipage}
}% \TABLE

\subsection{\bf Comments}

The $\Delta^{++}(1232)\Km$ mode is statistically significant even when
uncertainties associated with phases and other variables are taken into
account.  It is in fact comparable to the $p\Kstbar$ mode.  This
establishes the importance of the $W-$exchange mechanism in this charm
baryon decay.  Other channels in which $W-$exchange must contribute
include $\Sigma^+\phi$, $\Xiz\Kp$ and $\Xistz\Kp$ which have been
observed \cite{pdg:1999} with rates approximately half that of
$\Delta^{++}\Km$.

The observed sub channels in the $\Lc$ decay do not interfere
substantially, since sum of fractions is close to unity.

There is evidence for an increasingly negative polarisation of $\Lc$
baryons as a function of $p_t$.

There appears to be a dominant ($>50$\%) $NR$ component and the fit
is poor at low mass in the $\Km p$ system.  Many $\Lambda$ and $\Sigma$
states exist in this region, so the question arises how best to include
them in future analyses of this system without introducing too many
additional parameters.  Perhaps it is possible to introduce the (well
measured) published \cite{pdg:1978} $\Km p\ $ amplitudes and phases.  If
done properly, it is possible that much of the non resonant component
will thus be absorbed into this region.

%_______________________________________________________________________
\section{\bf Cabbibo Suppressed Decays}

Until a year ago only two experiments had observed Doubly Cabbibo 
Suppressed Decays (DCSD) of D mesons to $K\pi\pi$.  {\em No} baryons had
been observed to decay in even a singly Cabbibo suppressed (SCSD) mode.
Several results on Cabbibo Suppressed
Decays (SCSD and DCSD) have been presented at recent conferences.
These are updated here.

When normalised to the corresponding Cabbibo favoured mode we expect
branching ratios
\begin{eqnarray*}
 \Gamma(SCSD) \div \Gamma(CF)&=&\tan^2\theta_c
 ~\approx~5.1\times 10^{-2}  \\
 \Gamma(DCSD) \div \Gamma(CF)&=&\tan^4\theta_c
 ~\approx~2.6\times 10^{-3}
\end{eqnarray*}
Deviations from these ratios could be due to the influence of Final State
Interactions (FSI), hadronisation effects, interference between
different decay diagrams, {\bf or} higher order processes.

In the $\Dp\to\Km\pip\pip$ case (Cabbibo Favoured) destructive 
interference is {\em expected} between internal and external
spectator diagrams, so that it is anticipated that
  $$ {\Gamma(\Dp\to\Kp\pip\pim)\over\Gamma(\Dp\to\Km\pip\pip)}
  \approx{\tau_{\Dp}\over\tau_{\Dz}}\times\tan^4\theta_c $$
The experimental situation a year ago is summarised in table
\ref{cabibbo-yrago} where it is seen that this prediction is quite
well met by the data ($\tau_{\Dp}/\tau_{\Dz}\sim 2.5$).
\TABLE[ht]{%
\begin{minipage}[t]{.5\textwidth}
 \begin{center}
 \begin{tabular}{ l l c }
 \hbox{\bf\hss Ratio \hss}
 & Measured
 & Experiment \\
 & $\times\tan^{n}\theta_c$
   \footnote{Based on $\tan^2\theta_c~\approx~5.1\times 10^{-2}$
   and   $\tan^4\theta_c~\approx~2.6\times 10^{-3})$.  For singly
   Cabbibo suppressed decay, $n=2$ and for doubly suppressed $n=4$.}
 &                \\
 $\left({\Dp\to\Kp\pim\pip\over\Dp\to\Km\pip\pip}\right)        $  &
 $ 2.9\pm 0.6 $ &
  E687, E791
 \\
 $\left({\Dp\to\Km\Kp\Kp\over\Dp\to\Km\pip\pip}\right)          $  &
 $ < 0.6 $      &
  E687
 \\
 $\left({\Dz\to\Kp\pim\over\Dz\to\Km\pip}\right)                $  &
 $ 2.8\pm 1.0 $ &
  E791, CLEO2
 \\
 $\left({\Dz\to\Kp\pim\pim\pip\over\Dz\to\Km\pip\pim\pip}\right)$  &
 $ 1.0\pm 1.4 $ &
  E791, CLEO2
 \\
 \end{tabular}
 \end{center}
 \label{cabibbo-yrago}
 \caption{Summary of measurements of Cabbibo suppressed
          decays a year ago.}
\end{minipage}
}% \TABLE
More recent measurements of these ratios are summarised in table
\ref{cabbibo-recent}
\TABLE[ht]{%
\begin{minipage}[t]{.5\textwidth}
 \begin{center}
 \begin{tabular}{ l l c }
 \hbox{\bf\hss Ratio \hss}
 & Measured
 & Experiment \\
 & $\times\tan^{n}\theta_c$
 \\
 $\left({\Dp\to\Kp\pim\pip \over \Dp\to\Km\pip\pip}\right)$     &
 $ 2.4\pm 0.5 $                                                 &
  FOCUS $\sim$ 43\%
 \\
 $\left({\Dp\to\Km\Kp\Kp \over \Dp\to\Km\pip\pip}\right)$       &
 $ 0.62\pm 0.16 $                                               &
  FOCUS $\sim$ 43\%
 \\
 $\left({\Dsp\to\Kp\Kp\pim \over \Dsp\to\Km\Kp\pip}\right)$     &
 $ \sim 5 $                                                     &
  FOCUS $\sim$ 50\%
 \\
 $\left({\Xcp\to p\Km\pip \over \Xcp\to\Sigma^+\Km\pip}\right)$ &
 $ 4.0 \pm 1.3 $                                                &
  SELEX
 \\
 \end{tabular}
 \end{center}
 \caption{Recent {\bf preliminary} measurements of Cabbibo
          suppressed decays.}
 \label{cabbibo-recent}
\end{minipage}
}% \TABLE

The measurements by both CLEO and Focus of $\Dz$
hadronic decays which could be due either to Cabbibo suppressed decays
or to mixing.  These decays are discussed in more detail in 
\cite{sheldon:1999} in the $D$ Mixing talk in these proceedings.

\subsection{\bf New Evidence for $\Dp\hbox{ (and }\Dsp)\to\Kp\pip\pim$}

Using approximately 43\% of their data, FOCUS observes strong $\Dp$ and
$\Dsp$ signals shown in figure \ref{fig-kppmpp}.
These modes had already been observed by E791 \cite{Aitala:1997ht} and 
by E687 \cite{Frabetti:1995zi}.
% Their plots are shown in figures \ref{fig-791dcsd} and \ref{fig-687dcsd}
% respectively for comparison.
%
\FIGURE[ht]{%
  \epsfig{file=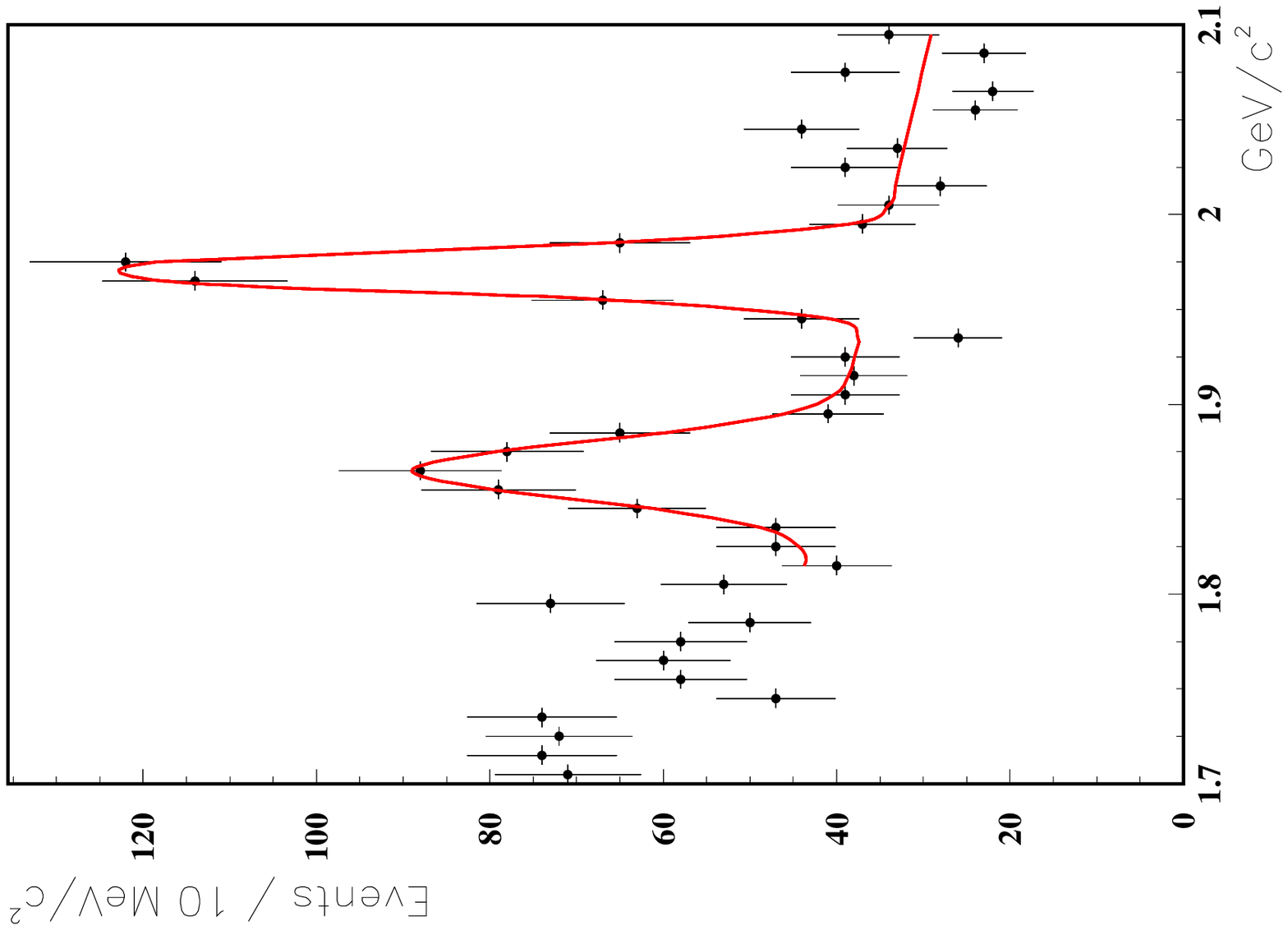,height=2.7in,angle=-90}
  \label{fig-kppmpp}
  \caption{FOCUS signal for doubly Cabbibo suppressed decays
   of $\Dp\to\Kp\pim\pip$ ($172\pm 28$ events).  The signal at higher
   mass is for $\Dsp$ decay to this final state ($224\pm 21$ Events.)}
}% \FIGURE
%
%\FIGURE[ht]{%
%  \epsfig{file=../Focus/link_791dcsd.eps,height=2.7in,angle=0}
%  \label{fig-791dcsd}
%  \caption{The E791 signal for doubly Cabbibo suppressed decays
%   of $\Dp\to\Kp\pip\pim$ ($59\pm 15$ events).}
%}% \FIGURE
%
%\FIGURE[ht]{%
%  \epsfxsize=2.7in
%  \epsffile{../Focus/link_687dcsd.eps}
%  \label{fig-687dcsd}
%  \caption{The E687 signal for doubly Cabbibo suppressed decays
%   of $\Dp\to\Kp\pip\pim$ ($21\pm 6.6$ events).}
%}% \FIGURE

The $\Dsp$ decay ($224\pm 21$ signal events) which is stronger in this 
mode than the $\Dp$ ($172\pm 28$ events) is SCSD while decay of $\Dp$ 
is DCSD.  E791 made a Dalitz plot analysis for the $\Dp$ and found 
roughly equal fractions of $\Kstz\pip$, $\Kp\rhoz$ and $NR$ 
$\Kp\pip\pim$.  FOCUS plans to make a similar analysis but with
about a factor six more events.

\subsection{\bf New DCSD and SCSD Observations}

The Focus experiment has preliminary evidence for several Cabbibo
suppressed decays \cite{Link:1998my}.  The decays
$\Dp(\Dsp)\to\Km\Kp\Kp$.
Their signals are shown in figure \ref{fig-3k}.
\FIGURE[ht]{%
  \epsfig{file=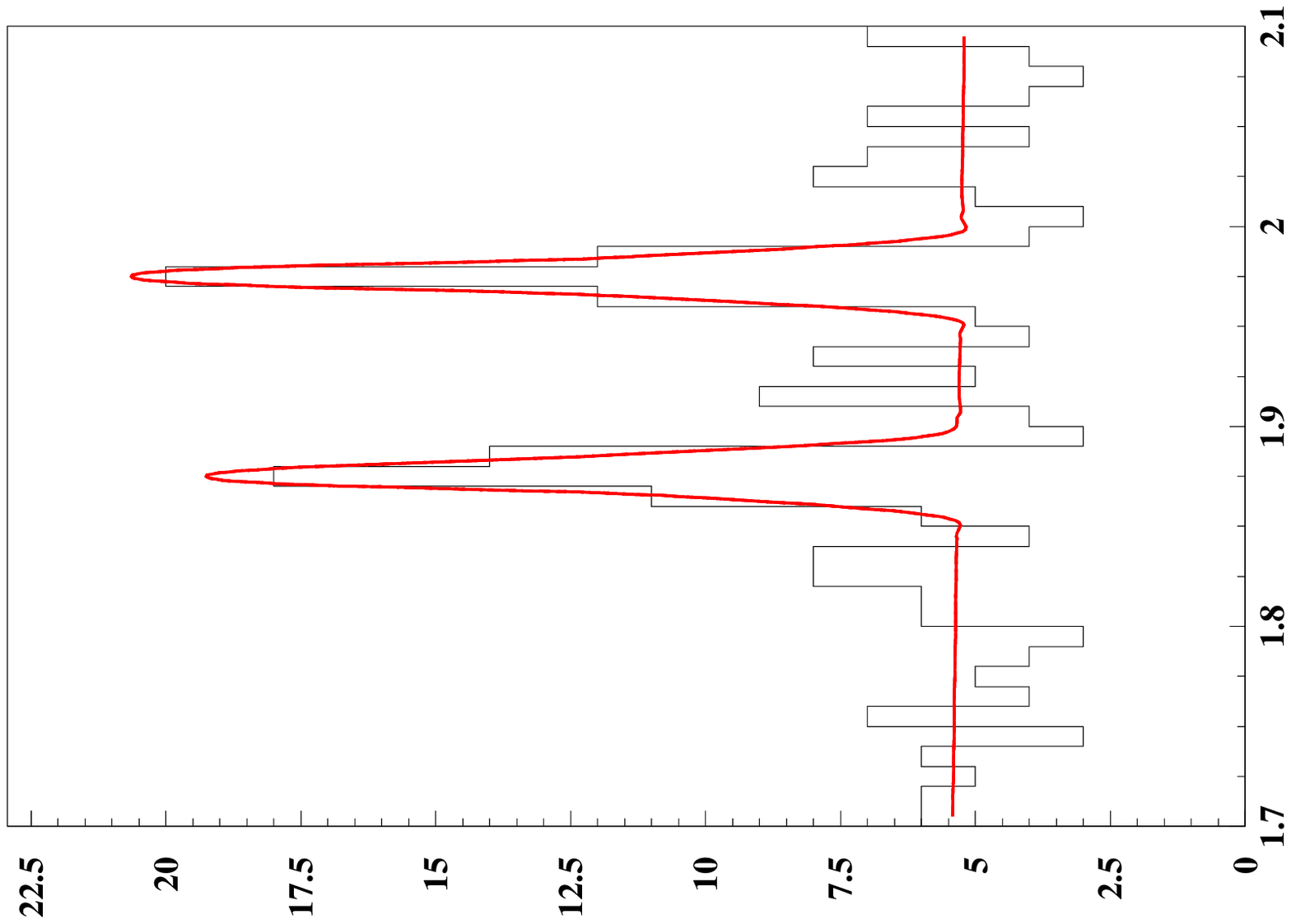,height=2.5in,angle=-90}
  \label{fig-3k}
  \caption{Preliminary evidence for $\Dp(\Dsp)\to\Km\Kp\Kp$ decays from
          Fermilab Experiment E831 (Focus).}
}% \FIGURE
The decay $\Dsp\to\Km\Kp\Kp$ is an example of SCSD, while the decay
$\Dp\to\Km\Kp\Kp$ is a DCSD.  In the latter case, this is the {\sl first}
observation of this mode.  There is no evidence for a $\phi\Kp$ sub
channel.  The interesting thing here is that there is no simple 
spectator process which leads to this mode.
\FIGURE[ht]{%
\begin{minipage}{0.40\textwidth}
  \centering
  \epsfig{file=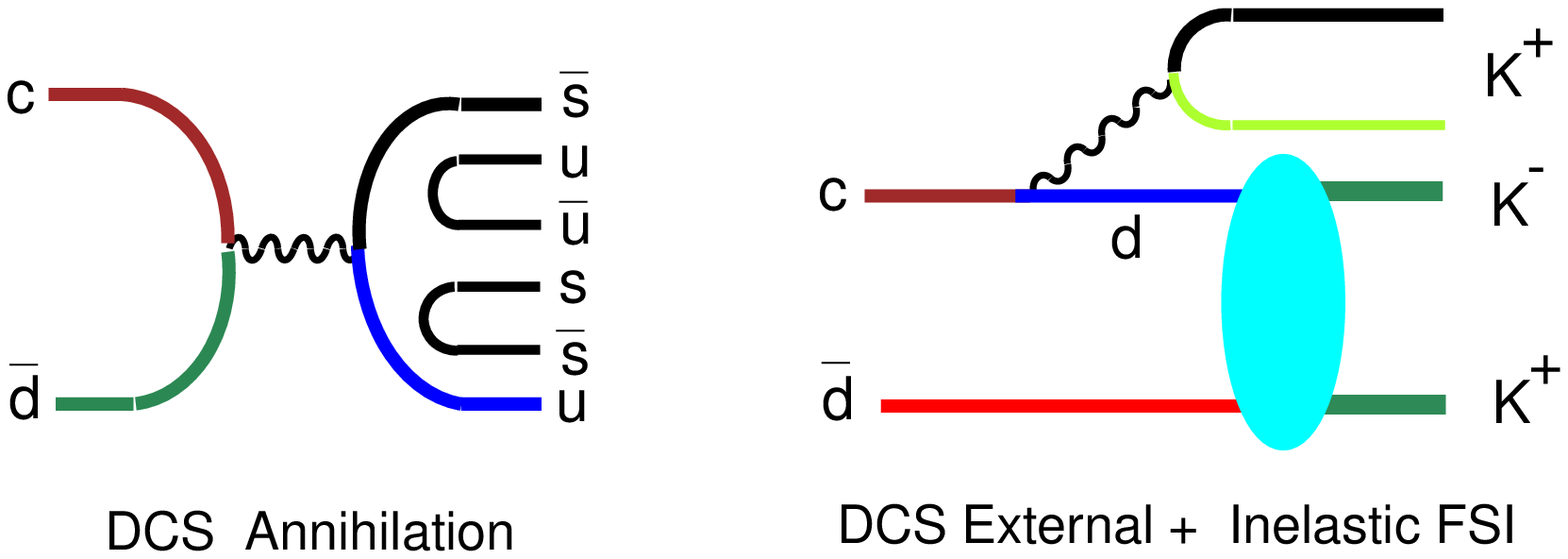,width=2.4in,angle=0}
  \caption{Decay diagrams for $\Dp\to\Km\Kp\Kp$.}
  \label{fig-kkk}
\end{minipage}
}% \FIGURE
Figure \ref{fig-kkk} shows the simplest ways for the decay to occur.
It appears that the decay rate could provide a way to estimate 
inelastic FSI from
   \[
    \begin{array}{ll}
     & {\Gamma(\Dp\to\pim\pip\Kp)\over \Gamma(\Dp\to\Km\Kp\Kp)}
     \\
     \approx & {\hbox{DCS External spectator} \over
               \hbox{DCS External spectator + Inel. FSI}}~~~.
    \end{array}
   \]

\subsection{\bf $\Dsp\to\pim\Kp\Kp$}
Also using a sample of 43\% of available data, FOCUS have
evidence (figure \ref{fig-dskpkppm}) for the decay $\Dsp\to\pim\Kp\Kp$.
This is the first observation of DCSD in $\Dsp$ decay.  The
preliminary rate quoted is
   \begin{eqnarray*}
   {\Dp\to\pim\Kp\Kp \over \Dp\to\Km\Kp\pip}
       &\approx& 1.5\%                        
        \approx  5\times\tan^4\theta_c
   \end{eqnarray*}
Lipkin has recently suggested that from $s\leftrightarrow d$
($U$-spin) symmetry the ratios
   \[
   \begin{array}{ccc}
   {\Dsp\to\pim\Kp\Kp  \over \Dsp\to\Km\Kp \pip}&\leftrightarrow&
   {\Dp \to\pim\Kp\pip \over \Dp \to\Km\pip\pip}      \\
   \\ 5\tan^4\theta_c&\leftrightarrow& 2.5\tan^4\theta_c
   \end{array}
   \]
should be found.
\FIGURE[ht]{%
  \epsfxsize=2.7in
  \epsfysize=2.0in
% \epsffile{../Focus/link_dskpkppm.eps}
  \epsffile{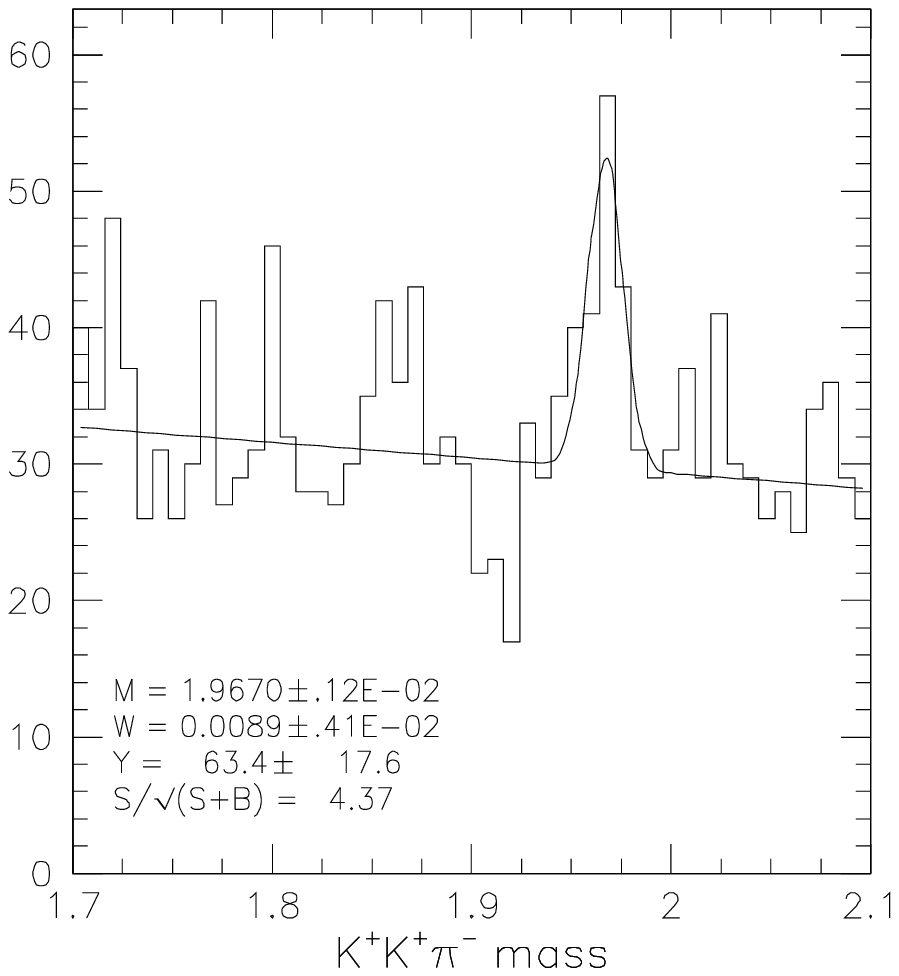}
  \caption{Evidence for $\Dp\to\Kp\Kp\pim$ from 43\% of available
           FOCUS data.}
  \label{fig-dskpkppm}
}% \FIGURE

\subsection{\bf $\Xcp\to p\Km\pip$}

This decay was seen first by the SELEX collaboration \cite{Jun:1999gn}
and is the first observation of a Cabbibo suppressed decay of a 
charmed baryon.  Their signal is shown in figure \ref{fig-xkppi}.
\FIGURE[ht]{%
\begin{minipage}{0.45\textwidth}
  \centering
  \epsfig{file=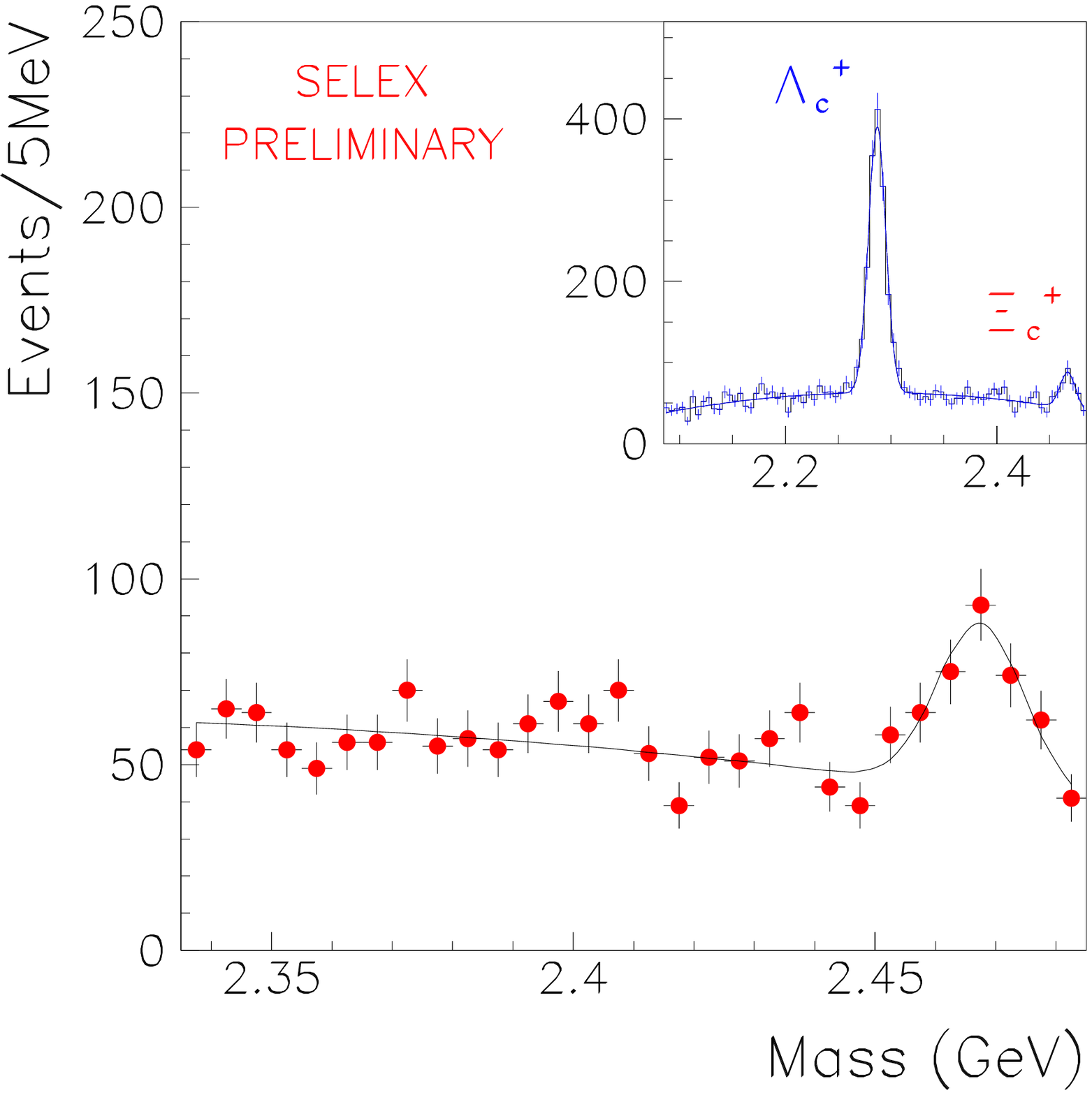,width=2.7in,height=2.0in,angle=0}
  \label{fig-xkppi}
  \caption{The decay $\Xcp\to p\Km\pip$ from the SELEX Collaboration.
   The signal observed is estimated to have $162\pm 19$ events and
   a significance (${S\over\sqrt{S+B}})$ of $7.0\pm 1.3$.}
\end{minipage}
}% \FIGURE

It has also been seen by FOCUS from 50\% of their available data as a 
signal with $73\pm 18$ events.

This is a SCSD and as seen in figure \ref{fig-xkppitree} it can be
related to the similar  decay mode for $\Sigma_c^+$.
\FIGURE[ht]{%
\begin{minipage}{0.40\textwidth}
  \centering
  \epsfig{file=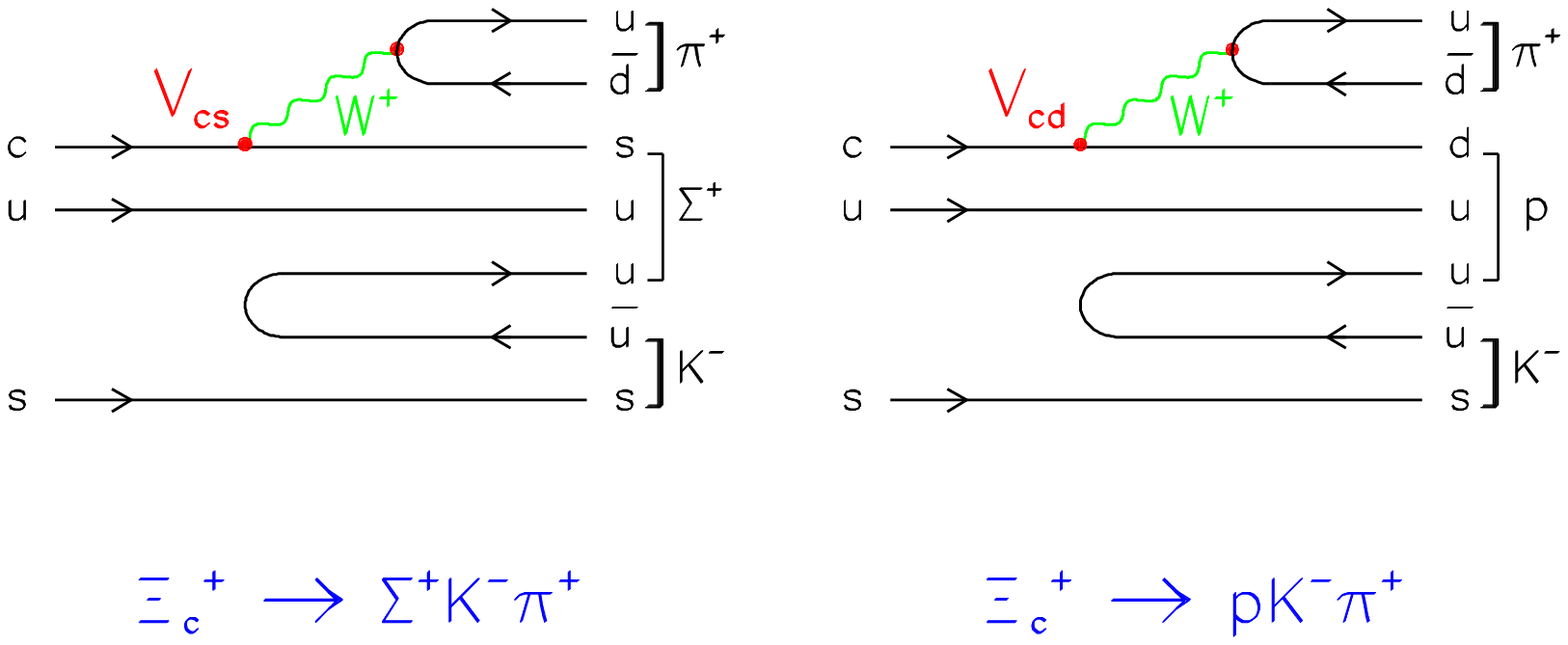,width=2.3in,angle=0}
  \label{fig-xkppitree}
  \caption{Baryon decays to $p\Km\pip$.}
\end{minipage}
}% \FIGURE
The SELEX collaboration have a preliminary measurement of the branching
ratio
$$ {{\cal B}(\Xcp\to p\Km\pip) \over {\cal B}(\Xcp\to\Sigma_c^+\Km\pip)}
 = 0.21\pm 0.07 \approx (4.0\pm 1.3)\tan^2\theta_c $$

\section{\bf Summary}

New information on hadronic decays of charmed mesons and baryons has
been obtained in the last year.
Some evidence for $\Dsp\to\rhoz(770)\pip$ and $\rhoz(1450)\pip$ has
been seen by E791.  This is either evidence for annihilation or for
inelastic FSI ($\Dsp\to\Km\Kp\pip$ followed by $\Km\Kp\to\pim\pip$
scattering.)  The latter is not unlikely it seems when one considers
that
  $\sigma(\pim p \to \Km\Kp\Lambda) \approx(0.05-0.10)
   \sigma(\pim p \to \pim\pip\Lambda)$
and that
  $\sigma(\pim p \to \Km\Kp\Lambda) \approx(0.05-0.10)
   \sigma(\pim p \to \pim\pip\Lambda)$
over a wide energy range.

E791's analysis of the $\Lc\to p\Km\pip$ decay shows the exchange diagram
for baryon decay to be more important than thought before.  Also, more
data on Cabbibo suppressed decays should eventually lead to more
quantitative information on FSI and interference effects.

Dalitz plot fits are getting more sophisticated, but a less subjective
way to judge which resonances to include is needed.  Partial wave
analyses may be made when more data are available perhaps
from E831 or the B factories.  Bose symmetrization makes it possible
in decays to final states involving identical particles such as
$\pim\pip\pip$ to determine phases as a function of mass without the
need to invoke a BW parametrization of an amplitude.  Also, future
analyses of $\Lc\to p\Km\pip$ decay should attempt a better description
of low mass $\Km p$ system.  Perhaps this would lead to a significantly
smaller non resonant mode.

New information on light quark states appears available when viewed from
the unusual perspective of $D$ decay.
Little is known of the light quark scalars and perhaps $D$ decays such as
$\Dp\to\Km\pip\pip$, $\Dp\to\Km\Kp\pip$, \etc. might shed further light
on these extremely important states.  E791 data has provided improved
parameters for $\fz(980)$ and $\fz(1370)$ and may have produced new
evidence for the $\fz(400-1200)$ (PDG nomenclature.)  This state could
be important to understanding the $\Delta I=\half$ rule or to models
of chiral symmetry breaking.
Inclusion of this state in the E791 analysis also has lead to a
satisfying picture in which both $\Dsp$ as well as $\Dp$ decays are
dominated by quasi two body modes.

So, there appears to be the need to shift the emphasis of studies of 
charmed hadronic decays from the investigation of decay dynamics to a 
better understanding of the light quark states to which they decay.  
Until this is achieved, it appears that further progress will remain 
frozen.  It could also be that we will eventually learn more about charm
spectroscopy from $B$ decays when the $B$ factories really start to
produce significant amounts of data.
CLEO has already shown how this
might work with the discovery of a new $D_1(j=1/2)$ state from
$B\to\Dstr\pi\pi$.

\bibliographystyle{JHEP}
\bibliography{proc}
\end{document}